\documentclass[acmsmall]{acmart}
\AtBeginDocument{%
  \providecommand\BibTeX{{%
    \normalfont B\kern-0.5em{\scshape i\kern-0.25em b}\kern-0.8em\TeX}}}

\setcopyright{acmlicensed}
\acmJournal{TOSEM}
\acmYear{2021}
\acmVolume{1}
\acmNumber{1}
\acmArticle{1}
\acmMonth{1}
\acmPrice{15.00}
\acmDOI{10.1145/3491.039}

\usepackage{tcolorbox}
\usepackage{url}
\usepackage{multirow}
\usepackage{tabulary} 
\usepackage[caption=false]{subfig} 
\usepackage{lipsum} 

\newcommand{\liwc}{\textsc{LIWC}}
\newcommand{\pin}{\textsc{IBM~PI}}
\newcommand{\pr}{\textsc{Personality Recognizer}}
\newcommand{\tp}{\textsc{TwitPersonality}}
\newcommand{\SE}{software engineering}
\newcommand{\asf}{Apache Software Foundation}
\newcommand{\fullcite}[1]{\citeauthor{#1}~\cite{#1}}
\newcommand{\code}[1]{{\small\colorbox{gray!10}{\texttt{#1}}}}
\newcommand{\tildex}[1]{$_{\widetilde{~}}$#1}
\newif\ifdiff
\newcommand{\revone}[1]{\ifdiff{\leavevmode\color{blue}{#1}}\else{#1}\fi}
\newcommand{\revonecaption}{\ifdiff\captionsetup{labelfont={color=blue},font={color=blue}}\else{}\fi}
\newcommand{\revonetable}{\ifdiff\color{blue}\else{}\fi}

\difffalse
\begin{document}

\title[Using Personality Detection Tools for Software Engineering Research]{Using Personality Detection Tools for Software Engineering Research: How Far Can We Go?}

\author{Fabio Calefato}
\email{fabio.calefato@uniba.it}
\orcid{0000-0003-2654-1588}
\author{Filippo Lanubile}
\email{filippo.lanubile@uniba.it}
\orcid{0000-0003-3373-7589}
\affiliation{%
  \institution{University of Bari}
  \streetaddress{Dipartimento di Informatica, via E. Orabona 4}
  \city{Bari}
  \country{Italy}
  \postcode{70125}
}

\renewcommand{\shortauthors}{Calefato and Lanubile}

\begin{abstract}
Assessing the personality of software engineers may help match individual traits with the characteristics of development activities such as code review and testing, as well as support managers in team composition. 
However, self-assessment questionnaires are not a practical solution for collecting multiple observations on a large scale. Instead, automatic personality detection,  while overcoming these limitations, is based on off-the-shelf solutions trained on non-technical corpora, which might not be readily applicable to technical domains like \SE. 
In this paper, we \revone{first} assess the performance of general-purpose personality detection tools when applied to a technical corpus of developers' emails retrieved from the public archives of the \asf.
We observe a general low accuracy of predictions and an overall disagreement among the tools.
\revone{Second}, we replicate two previous research studies in \SE\ by replacing the personality detection tool \revone{used to infer developers' personalities from pull-request discussions and emails. We} observe that the original results are not confirmed, i.e., changing the tool used in the original study leads to diverging conclusions.
Our results suggest a need for personality detection tools specially targeted for the software engineering domain. 
\end{abstract}



\maketitle

\section{Introduction}

Software engineers' personality has been a subject of interest for researchers since the 1970s when it was first hypothesized that personality traits might influence how developers interact~\cite{weinberg1971psychologyse,shneiderman1980psychologyse}. 
\fullcite{cruz2015slr} identified 90 studies published between 1970 and 2010, most of which published after 2002; \fullcite{barroso2017slr} found 21 studies, published between 2003 and 2016, studying the effect of personality on professional developers. 
The main reasons for this growing interest in personality-focused research lie in the many practical implications existing at both individual and team level.
For example, previous studies on the personality of developers involved in \revone{agile development} have revealed a positive association of conscientiousness (i.e., being organized, dependable) and openness to experience with their \revone{pair programming} performance~\cite{salleh2010icsemotivation,salleh2010esemmotivation}\revone{;}
\revone{in addition, evidence suggests that testers are significantly higher on conscientiousness than other software development practitioners \cite{kanij2015motivation} and that managers are more extroverted \cite{smith2016personalityse}.} 
Regarding performance, teams whose members are more extroverted have been found to release higher-quality software products~\cite{acuna2009motivation,acuna2015motivation}.
\fullcite{licorish2009supporting} have built a prototype to assist project managers in agile team formation by providing lightweight support for personality 
assessment.
\revone{However, these studies barely scratched the surface of the potential scenarios of applying personality to \SE, given its profound socio-technical nature~\cite{mauerer2021stc}. 
}

Most of the previous research on personality has been conducted using self-assessment questionnaires.
Albeit reliable,  detecting the personality through psychometric questionnaires has some drawbacks, such as low return rates---especially in the \SE\ domain~\cite{smith2013sesurveys}---and
the limited number of occasions (typically one) to perform data collection~\cite{yarkoni2010liwc}. The drawbacks of self-assessment questionnaires can be overcome by computational personality detection, which is the task of automatically inferring personality traits from conversation transcripts and written text~\cite{argamon2005first,vinciarelli2014apr,farnadi2016cpd}.

With the proliferation of collaborative development environments such as GitHub, social media like Twitter, and communication platforms like Slack, developers' discussions have become easily accessible.
Thanks to the availability of such a wealth of communication traces, \SE\ researchers have begun developing solutions for automatically detecting developers' personalities from written text, thus making it possible to study on a large scale whether and how specific aspects might influence the outcome of development activities.
For example, previous research has relied on automatic tools to investigate how personality varies with the level of developers' contribution and prominence in the Apache~\cite{rigby2007liwc-asf}, Stack Overflow~\cite{bazelli2013liwc-so}, and GitHub~\cite{rastogi2016liwc-gh} ecosystems; 
other studies have focused on clustering similar personality profiles among the developers of the Eclipse~\cite{paruma2016ibmpi,licorish2015personalitygsd} and Apache~\cite{calefato2019apache} projects. 

The research on computational personality detection has resulted in the release of several general-purpose prediction tools and models (e.g.,~\cite{mairesse2007pr,pennebaker2015liwc,liu2017c2w2s4pt,majumder2017deeplearning,carducci2018twitpers}).
However, while there has been prior work that assessed the psychometric validity of questionnaires showing good correlations among the instruments (e.g., \cite{schmitt2007bfi}), when it comes to studies on automatic personality detection we found no previous assessment of general-purpose tool performance used as off-the-shelf components to analyze the communication traces left in a technical domain such as \SE. 
Therefore, to fill this gap, in this paper we investigate the problem of using computational personality detection tools trained on non-technical domains for \SE\ research.
In particular, we first ask:
\begin{itemize}
     \item \noindent \textbf{RQ1} -- \textit{How do off-the-shelf personality detection tools perform in the \SE\ domain?}
\end{itemize}

The findings of this study show that when general-purpose personality detection tools are used off-the-shelf for \SE\ research, their performance is far from acceptable as (i) neither they agree with self-reported ratings nor with each other; (ii) their prediction accuracy is lower compared to the results from prior work in non-technical domains.

The low agreement and accuracy rates of personality predictions suggest that  the conclusions based on the application of these tools in the \SE\ domain might be affected by the choice of one specific tool over the others.
Therefore, to understand whether using different personality detection tools affects the validity of the results when employed in the same \SE\ study, we ask:
\begin{itemize}
    \item \noindent  \textbf{RQ2} -- \textit{How does the choice of a personality detection tool affect the validity of previous results in \SE\ research?}
\end{itemize}

We conduct the replications of two former studies and---after changing the tool used therein--- find that the choice of a specific personality detection tool does affect the validity of the previously published results in \SE\ research as we generally fail to reproduce them. 

From a research perspective, our study makes a first attempt at benchmarking the performance of the  tools available for personality detection from technical text.
The study advances the state of the art on computational personality detection for \SE\ research by suggesting that, to cope with a domain-specific lexicon, we need to develop \SE-specific tools given that the existing solutions cannot be fine-tuned because the prediction models are generally not retrainable.

From a practical perspective, this study furthers our understanding of the current limitations when reusing computational personality detection tools as off-the-shelf components. 
We show that tools underperform in presence of technical text and warn that the choice of specific personality detection tools may lead to contradictory results.
Another practical contribution is the replication package that we share with the research community, including both the scripts for automatically re-executing the entire experimental workflow and an anonymized gold standard consisting of an email corpus matched with developers' self-reported personality scores.

The remainder of this paper is organized as follows. 
In Sect.~\ref{sec:background}, we illustrate the study background.
Sect.~\ref{sec:res-framework} presents the two-phase research framework followed to carry out our work.
In Sect.~\ref{sec:agreement-analsysis}, we study the performance of the selected personality prediction tools.
In Sect.~\ref{sec:replicated-studies}, we perform the replication of two previously published, \SE\ studies. 
The results are discussed in Sect.~\ref{sec:discussion}.
Finally, we present the related work in Sect.~\ref{sec:rel-work} and conclude in Sect.~\ref{sec:conclusions}.


\section{Background}
\label{sec:background}

In Sect.~\ref{sec:personality}, we first provide an overview of the fundamental concepts and theories related to personality. 
Then, in Sect.~\ref{sec:personality-detection-instruments} we review some of  
the instruments used for personality measurement.
Finally, in Sect.~\ref{sec:personality-in-se}, we review  prior work that focused on developers' personalities in the \SE\ field.

\subsection{Personality Theories}
\label{sec:personality}

Personality is defined by psychologists as the set of all the behavioral, temperamental, emotional, and mental attributes that characterize a unique individual~\cite{ryckman2012theories}. 
Personality has been conceptualized from a variety of theoretical perspectives and at various levels of abstractions. 
One level that has been often studied is \textit{personality traits}~\cite{john1999big5traits}.
Given the complexity of its nature, psychologists have developed several taxonomies of traits, that is, descriptive models of personality useful for organizing, distinguishing, and summarizing the major individual differences among the numerous existing in human beings.

Many theories of personality traits have been proposed since the 1930s. 
Such theories often disagreed regarding the number of traits and their nature~\cite{goldberg1993traits}.
However, after decades of research and the compelling amount of empirical evidence collected, since the 1970s a general consensus was achieved on the validity of a taxonomy of five orthogonal personality traits, called the \textit{Big Five}. 
A brief description of the five factors is reported in Table~\ref{tab:b5}.
The name was first proposed by Goldberg~\cite{goldberg1981bigfive} to emphasize that five dimensions are sufficient to capture the main dispositional characteristics and high-level differences of individuals.
These five personality traits have been obtained through repeated studies by applying factor analyses to various lists of trait adjectives used in self-descriptions and self-rating questionnaires for personality assessment. 
These studies are based on the \textit{lexical hypothesis}~\cite{allport1936lexical-hypothesis}, a psycholinguistic conjecture according to which the most important individual characteristics and differences in personality have been encoded over time as words in the natural language, and the more important a characteristic or difference, the more likely it is for individuals to express it using associated words~\cite{john1999big5traits}.

\revone{Several independent studies (see \fullcite{digman1990ffm} for a compendium) performed repeated factor analyses on questionnaires based on different personality taxonomies only to find consistent evidence of the existence of a latent personality structure consisting of five main factors.}
\revone{In other words, t}he extracted models showed minor differences at the higher level---i.e., the traits, albeit labeled differently, could be easily mapped onto each other~\cite{goldberg1993traits}. 
Hence, trait psychologists  combined these findings on the general ubiquity of five factors across various instruments  with the results from the studies on the lexical hypothesis to argue that any personality traits model had to encompass at some level the same Big Five dimensions~\cite{goldberg1981bigfive}.

Albeit the Big Five is considered a synonym with the \textit{Five-Factor Model} (FFM)~\cite{costa1985neo-pi,mccrae1987ffm-validation}, the two models are slightly different.
Big Five is a term used to refer in general to personality frameworks consisting of five high-level dimensions and the Big Five model only describes the five traits at a broad level. 
Instead, the FFM is a personality framework that further refines the five high-level traits into multiple lower-level facets. 
For the sake of simplicity, from now on we will use Big Five and FFM interchangeably.

\begin{table}[t]
\caption{Overview of the Big Five traits and how their low vs. high levels characterize individuals. The traits are often referred to by the mnemonic OCEAN (adapted from \fullcite{plotnik2013intro-psychology}).}
\label{tab:b5}
\small
\begin{tabular}{cll}
\hline
\multirow{2}{*}{\textbf{Factor}}  & \multicolumn{2}{c}{\textbf{Description}}                                                                                                                                                                                      \\ \cline{2-3} 
                                  & \multicolumn{1}{c|}{\textbf{Low}}                                                                                           & \multicolumn{1}{c}{\textbf{High}}                                                               \\ \hline
\multirow{2}{*}{Openness}         & \multicolumn{2}{l}{\begin{tabular}[c]{@{}l@{}}Refers to the extent to which a person is open to experiencing a variety \\ of activities, proactively seeking and appreciating unfamiliar experiences\end{tabular}}            \\ \cline{2-3} 
                                  & \multicolumn{1}{l|}{\textit{Is conservative and close-minded}}                                                              & \textit{Is open to novel experiences}                                                           \\ \hline
\multirow{2}{*}{Conscientiousness} & \multicolumn{2}{l}{\begin{tabular}[c]{@{}l@{}}Refers to the tendency to plan in advance, act in an organized or thoughtful \\ way as well as the degree of persistence and motivation in goal-directed behavior\end{tabular}} \\ \cline{2-3} 
                                  & \multicolumn{1}{l|}{\textit{Is impulsive and careless}}                                                                     & \textit{Is responsible and dependable}                                                          \\ \hline
\multirow{2}{*}{Extroversion}     & \multicolumn{2}{l}{\begin{tabular}[c]{@{}l@{}}Refers to the tendency to seek stimulation in the company of others, thus\\ assessing a person's amount of interpersonal interaction and activity level\end{tabular}}           \\ \cline{2-3} 
                                  & \multicolumn{1}{l|}{\textit{Is reserved and solitary}}                                                                      & \textit{Is outgoing and decisive}                                                               \\ \hline
\multirow{2}{*}{Agreeableness}    & \multicolumn{2}{l}{\begin{tabular}[c]{@{}l@{}}Refers to a person's tendency to be cooperative with others and \\ compassionate about their thoughts, feelings, and actions\end{tabular}}                                      \\ \cline{2-3} 
                                  & \multicolumn{1}{l|}{\textit{Is unfriendly and cold}}                                                                        & \textit{Is warm and good-natured}                                                               \\ \hline
\multirow{2}{*}{Neuroticism}      & \multicolumn{2}{l}{\begin{tabular}[c]{@{}l@{}}Refers to the extent to which one lacks emotional stability and is prone \\ to psychological distress, anxiety, excessive cravings, or urges\end{tabular}}                      \\ \cline{2-3} 
                                  & \multicolumn{1}{l|}{\textit{Is stable and calm}}                                                           & \textit{Is nervous and emotionally unstable}                                                                     \\ \hline
\end{tabular}
\end{table}

\subsection{Instruments for Personality Detection}
\label{sec:personality-detection-instruments}

Psychometrics is the development of measurement instruments and the assessment of whether these instruments are reliable and valid forms of measurement~\cite{ginty2013psychometrics}.

\subsubsection*{Self-rating Questionnaires}
Personality traits are usually assessed using self-assessment questionnaires, which present a variable number of items (up to hundreds) that describe common situations and behaviors. 
Subjects take these self-reporting tests by rating on a Likert scale the extent to which an item applies to them, where each item is either positively or negatively associated with a specific trait. 
Finally, a numeric score is computed for each trait by aggregating (e.g., summing) all the values assigned to its related answers.

There are several instruments for the self-assessment of personality, such as the   \textit{Myers-Briggs Type Indicator} (MBTI)~\cite{myers1998mbti} and the \textit{Keirsey Temperament Sorter} (KTS)~\cite{keirsey1998kts}, based on Jung's theories. 
However, their psychometric validity has been questioned over the years~\cite{boyle1995reliability,abramson2010reliability}


The most popular instruments for measuring the Big Five traits are the \textit{NEO-PI}~\cite{costa1985neo-pi} and the \textit{NEO-PI-R}~\cite{costa2008neo-pi-r}. 
McCrae~\cite{mccrae2001culture,mccrae2002culture} has found NEO-PI-R to be reliable even after translating and administrating it across 36 countries, also showing that it is possible to use trait mean values to capture systematic differences.
Another questionnaire intended to measure the five dimensions of personality is the \textit{Big Five Inventory} (BFI) used by \fullcite{schmitt2007bfi} in a large study on 56 nations. 
The results showed a robust five-factor structure across geographical and cultural regions as well as a high cross-instrument correlation with the NEO-PI-R scales. 

Because all the instruments above are proprietary, psychologists have developed and validated the \textit{International Personality Item Pool} (IPIP) and its follow-up IPIP-NEO (\textit{International Personality Item Pool 
Representation of the NEO PI-R}), two alternative and open Big Five inventories freely available to researchers \cite{goldberg2006ipip}.

In summary, given the evidence of the validity gained by the Big Five inventories in general and the lack of psychometric reliability of the other instruments, in this work, we focus only on the FFM and the related instruments. 

\subsubsection*{Computational Personality Detection from Text}
Self-report inventories are the most popular psychometric instruments to assess personality because of their validity and ease of use.
However, in addition to its semantic content, written text is also capable of conveying information about the writer, such as cues to individual personality.
Psychologists have been able to identify correlations between specific linguistic markers and personality traits~\cite{pennebaker1999linguistic}.
Computational personality detection~\cite{farnadi2016cpd}, also referred to as automatic personality recognition~\cite{vinciarelli2014apr} is the task of inferring   \revone{people's personality from their digital footprints, such as  social media content (e.g., videos, pictures, likes), and written text (e.g., conversation transcripts, blog posts,  emails)}.
\revone{A machine-learning algorithm uses features extracted from the analyzed content as cues to predict personality.
The features and how they are associated with the traits vary depending on the type of corpora analyzed; for instance, audio recordings allow the use of acoustic cues, such as voice inflection, which are lacking in textual corpora where instead lexical features, such as the count of positive vs. negative words, part-of-speech tags, and \textit{n}-grams, can be leveraged as personality markers.
Further details on how computational personality detection works are provided later in Section~\ref{sec:tools}, where the selected tools are introduced.}

To date, there is a limited yet steadily growing amount of work on the automatic detection of personality~\cite{kaushal2018emerging}. 
In particular, thanks to the recent advances in machine learning, recently developed tools (e.g., \cite{liu2017c2w2s4pt,carducci2018twitpers}) are leveraging deep-learning techniques for processing several cues extracted from large text corpora.
Tools can be grouped into top-down and bottom-up solutions~\cite{celli2013workshop}. 
Top-down (or closed-vocabulary) solutions rely on external resources (e.g., psycholinguistic databases) and text is processed through such pre-determined dictionaries that specify meaningful word categories associated \textit{a priori} with personality traits.
Instead, bottom-up (or open-vocabulary)~\cite{schwartz2013openvocabulary} do not specify in advance the relationship between features and allow linguistic cues (i.e., meaningful words and phrases) associated with personality traits to emerge from data. 

In Sect.~\ref{sec:tools}, we review in detail some of the tools available as off-the-shelf solutions to automatically recognize the Big Five personality traits from text. 

\subsubsection{Personality Datasets}
\label{sec:personality-datasets}

Because automatic personality recognition approaches are inherently data-driven, the availability of experimental datasets plays a crucial role.
According to~\fullcite{novikov2021survey}, more than 40\% of studies on personality involve the  collection of new datasets that remain private. The remaining studies rely on a few shared and reusable datasets.
Here we briefly review those based on textual documents.

The essay dataset~\cite{pennebaker1999linguistic} consists of nearly 2,500 
essays (i.e., unedited pieces of text) written in a controlled setting by students who had also taken the BFI personality test. 
Originally used to validate the \liwc\ tool (see Sect.~\ref{sec:tools}), it is one of the first corpora utilized in personality prediction research~\cite{argamon2005first,mairesse2007pr} and is still very much in use~\cite{mehta2020essay,salminen2020essay}.  

Given the growing amount of digital traces left by users in social networks, it is not surprising that most of the personality datasets collect documents from social media (e.g., ~\cite{liu2016socialmedia,hall2017socialmedia,ramos2018socialmedia}).
The largest dataset used in personality prediction research is myPersonality~\cite{stillwell2004mypersonality}, which contains data of Facebook users who filled in a personality questionnaire. 
The anonymized dataset has been freely shared with researchers for non-commercial academic purposes until it was retired in 2018.
Another example of personality-annotated datasets of posts from social
networks is the PAN-AP-2015 corpus~\cite{rangel2015pantask}, consisting of Twitter posts in English, Spanish, Italian, and Dutch from users who also took the BFI test.

To the best of our knowledge, there is no personality-annotated dataset of text documents (e.g., emails, issue reports, commit messages) collected from technical domains such as software engineering.

\subsection{Personality Detection in Software Engineering}
\label{sec:personality-in-se}

Early in the development of the \SE\ field, it was recognized that, in addition to technological factors, researchers also had to consider the humans involved in the development process~\cite{weinberg1971psychologyse,shneiderman1980psychologyse}. 
\fullcite{lenberg2015bse} have proposed the term Behavioral Software Engineering to refer to the interdisciplinary study of cognitive, behavioral, and social aspects of \SE\ as performed by individuals and groups.


In the following, we review the most relevant previous studies on personality in \SE.
We restrict our review to studies published since 2016---for earlier studies, refer to the SLRs reported in~\cite{cruz2015slr,barroso2017slr}.
Also, given the compelling amount of evidence on its validity, we focus our review only on studies that leveraged the Big Five model.

\fullcite{kosti2016personalityse} conducted a study using a clustering technique, which identified four archetypal personality profiles characterized by the levels of \textit{extraversion} and \textit{conscientiousness}.
\fullcite{mellblom2019personalityse} analyzed the response from 47  participants in a survey aimed at revealing the relationship between specific personality traits and burnout in professional software developers.
Through regression analysis, they uncovered a strong link between \textit{neuroticism} and burnout.
\fullcite{mendes2021decisionmaking} surveyed 63 Brazilian developers and found that the \textit{agreeableness} trait is significantly associated with the variation in the decision-making style. 
\fullcite{smith2016personalityse} analyzed the characteristics of professional developers’ personalities based on their roles. They found managers to be more \textit{conscientious} and \textit{extroverted}, and agile developers to be more \textit{neurotic} and \textit{extroverted}.
\fullcite{akarsu2019personalityse} studied the personality traits by administering the BFI to 18 agile teams.
They found that high levels of \textit{agreeableness} and \textit{conscientiousness} were common in most teams. 
Moreover, they observed a low level of \textit{extraversion} in isolated teams that had fewer contacts with customers. 
In~\cite{vishnubhotla2020personalityse},~\citeauthor{vishnubhotla2020personalityse}  investigated the association between the Big Five traits and the factors related to team climate within eight agile teams.
Through regression analysis, they found that \textit{openness} has a statistically significant positive correlation with support for innovation; also \textit{agreeableness} is positively correlated with the overall team climate.
\revone{Finally, an interesting attempt was made by~\fullcite{yilmaz2017personalityse} who developed a psychometric questionnaire, based on the BFI and specifically adapted it to the \SE\ domain, to explore how practitioners' personality traits are associated with effective software teams.
The results indicated that effective teams are characterized by low \textit{neuroticism} and high levels of \textit{agreeableness}, \textit{extraversion}, and \textit{conscientiousness}. }

All the studies reviewed above rely on questionnaires for psychometric assessment.
Yet, several studies also rely on the automatic recognition of developers' personality traits from text.
In the Related Work (Sect.~\ref{sec:rel-work}), we provide an in-depth review of such studies.


\section{Research Framework}
\label{sec:res-framework}

\begin{figure}[t]
\centerline{
\includegraphics[width=0.65\textwidth]{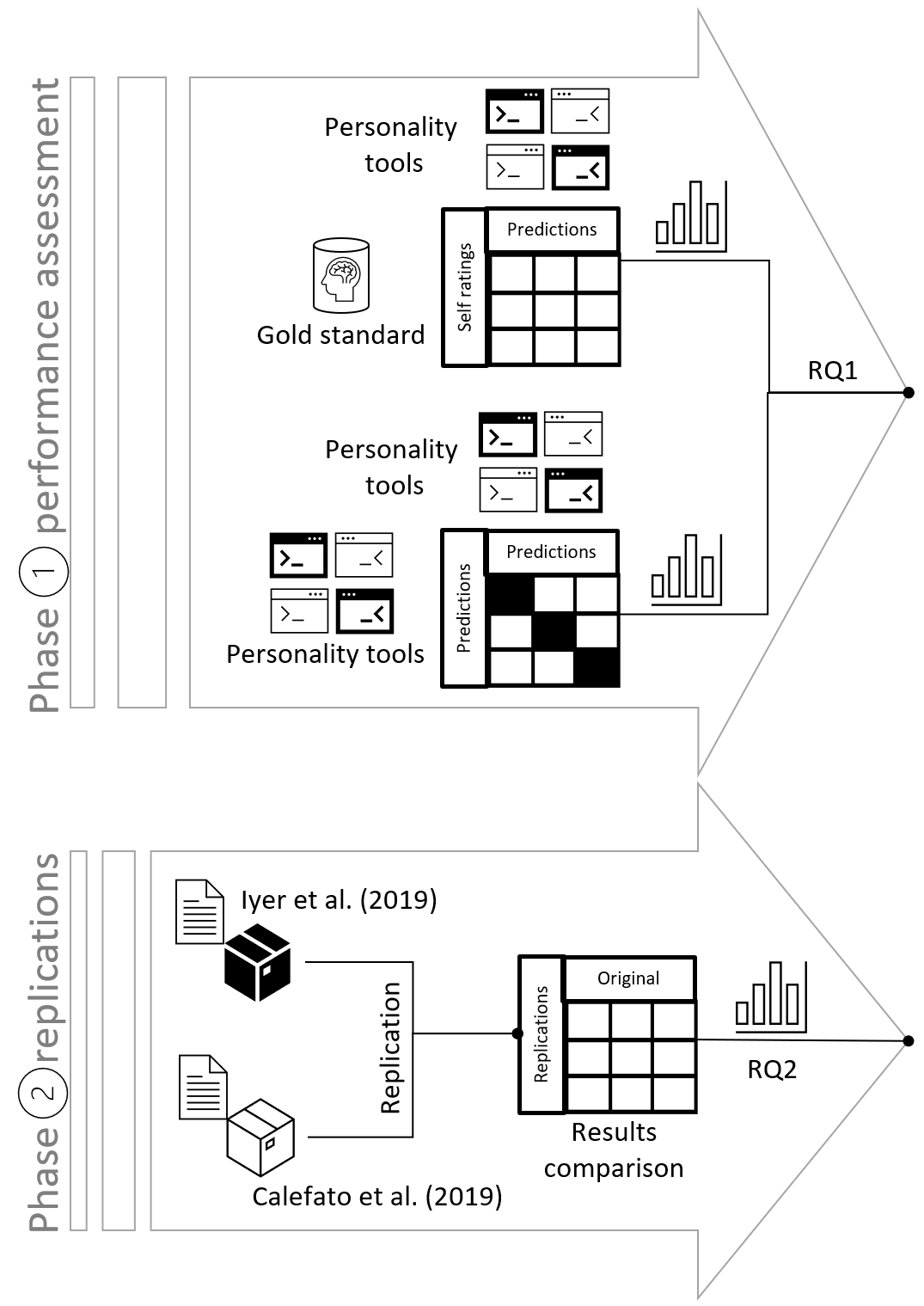}}
\caption{Overview of the research framework.}
\label{fig:res-workflow}
\end{figure}

In this work, we follow a two-phase research framework, as depicted in Fig.~\ref{fig:res-workflow}.

In \textit{Phase 1}, we assess the performance of personality detection tools for \SE. 
Accordingly, to answer RQ1, we first select four tools built upon the Big Five model.
Then, we set the ground truth by collecting the responses to a self-assessment personality questionnaire from 50 \asf\ (ASF) developers.  
Contextually, we build a dataset of over 1,500 emails written by the ASF developers. 
We apply the selected tools to the email dataset to extract the personality scores and compare their predictions against the self-ratings.
Finally, we complement the previous analysis by  assessing the agreement between the selected tools. 
As such, we make pairwise comparisons between the personality scores obtained by running the tools on the email corpus.

In \textit{Phase 2}, we study whether the choice of one personality detection tool over the others affects the validity of results from software engineering studies. 
Accordingly, to answer RQ2, we select two \revone{recent, large-scale studies, respectively  by \fullcite{iyer2019github} and \fullcite{calefato2019apache}} 
\revone{We choose to replicate these two studies because they both} used the same tool \revone{(i.e., IBM Personality Insights)} to extract developers' personality profiles from different datasets of technical content---respectively, \revone{pull-request discussions} obtained from GitHub and emails retrieved from the ASF public archives. 
\revone{Finally}, both studies provide a replication package, which we adapt to replicate them using another personality detection tool\revone{; the availability of complete replication packages allows us to minimize the risk of errors in executing the replications.}

\revone{The two-phase framework is inspired by the work of \fullcite{jongeling2017negative}, who conducted a study on the use of sentiment analysis tools for software engineering research and arranged their research questions sequentially so that, after observing disagreement among tool predictions, they could  explore the effects on conclusion validity in prior work replications after switching tools.}

We provide a complete replication package for the work presented here, which allows other researchers to re-execute all the steps in the research workflow. 
To reinforce the replicability, we provide a script that fully automates the whole experiment pipeline. Further details and instructions are available in Appendix~\ref{appendix:rep-package}.

                            
\section{Phase 1 -- Assessment of Prediction Performance}
\label{sec:agreement-analsysis}

This section is structured as follows. In Sect.~\ref{sec:tools}, we describe the process followed to select the sample of tools for computational personality detection from text; we also provide details about their features, configuration, and output. 
In Sect.~\ref{sec:goldstandard}, we describe how we built a gold standard by administering a self-assessment personality questionnaire to software developers.
In Sect.~\ref{sec:dataset}, we illustrate the process followed to build the dataset of emails written by the same subjects who answered the personality questionnaire.
Finally, In Sections~\ref{sec:metrics}~and~\ref{sec:ph1-results}, respectively, we describe the evaluation metrics and report the results to answer RQ1.

\begin{table}[t]
\caption{The selected tools for computational personality detection on a continuous numerical scale. In column \textit{Solution}, TD stands for Top-Down (closed vocabulary), BU for Bottom-Up (open vocabulary). The column \textit{Scale points} indicates how the output is measured.}
\label{tab:tools}
\tiny
\begin{tabular}{llcllllc}
\hline
\textbf{Tool}                                                    & \textbf{License}                                             & \multicolumn{1}{l}{\textbf{Approach}} & \textbf{Technique}                                                   & \textbf{Features}                                                         & \textbf{Dataset}                                                         & \textbf{\begin{tabular}[c]{@{}l@{}}Validation \\ (ground truth)\end{tabular}} & \textbf{\begin{tabular}[c]{@{}c@{}}Scale\\ points\end{tabular}}    \\ \hline
LIWC                                                             & Commercial                                                   & TD                                    & \begin{tabular}[c]{@{}l@{}}Word category \\ frequencies\end{tabular} & \begin{tabular}[c]{@{}l@{}}Closed\\ vocabulary\end{tabular}               & 2,479 essays                                                             & BFI                                                                           & \begin{tabular}[c]{@{}c@{}}word-trait\\ correlations*\end{tabular} \\ \hline
IBM PI                                                           & Commercial                                                   & BU                                    & \begin{tabular}[c]{@{}l@{}}Unspecified\\ ML technique\end{tabular}   & \begin{tabular}[c]{@{}l@{}}Open vocabulary,\\ word embedding\end{tabular} & 1,550-2,000 tweets                                                       & \begin{tabular}[c]{@{}l@{}}Unspecified Big 5\\ questionnaire\end{tabular}     & 5                                                                  \\ \hline
\begin{tabular}[c]{@{}l@{}}Personality\\ Recognizer\end{tabular} & \begin{tabular}[c]{@{}l@{}}Research\\ prototype\end{tabular} & TD                                    & \begin{tabular}[c]{@{}l@{}}Regression \\ models\end{tabular}         & LIWC, MRC                                                                 & 2,479 essays                                                             & LIWC dataset                                                                  & 7$^\diamond$                                                       \\ \hline
TwitPersonality                                                  & Apache 2.0                                                   & BU                                    & SVM                                                                  & Word embedding                                                            & \begin{tabular}[c]{@{}l@{}}18,473 tweets,\\  9,913 FB posts\end{tabular} & \begin{tabular}[c]{@{}l@{}}BFI, \\ myPersonality dataset\end{tabular}         & 5                                                                  \\ \hline
\multicolumn{8}{l}{\begin{tabular}[c]{@{}l@{}}* Transformed into trait scores using the formulae reported in~\fullcite{yarkoni2010liwc} and scaled into the range $1-5$.\\ $^\diamond$ Rescaled into the range $1-5$\end{tabular}}                                            
\end{tabular}
\end{table}

\subsection{Tools Selection}
\label{sec:tools}

In this section, we review in detail the personality detection tools selected for the analysis of prediction accuracy and agreement.

To build a list of candidate tools, we started from those listed in recent studies~\cite{calefato2019apache,mehta2019recenttrends} containing reviews of solutions for automatic personality recognition from text, including both commercial tools and research prototypes. 
Then, from the resource identified, we sought more candidates using a snowball method.
At the end of the search process, we identified \revone{19} candidates.\revone{\footnote{The complete list of the tools identified along with the exclusion criteria is available at \url{https://doi.org/10.6084/m9.figshare.15086391}.}}
From these, we filtered out those for which the  tool has not been shared or made available as an off-the-shelf solution for testing the model on other datasets. 
Tools vary largely in terms of the type of prediction task.
According to Schwartz et al.~\cite{schwartz2013openvocabulary}, prediction on a continuous numeric scale (i.e., traits are measured on a given numeric range) is a more appropriate task for studies on automatic personality recognition.
Therefore, we also filtered out those tools that intend personality trait prediction as a binary (e.g., yes/no) or multi-class (e.g., low/medium/high) task instead of measuring the outcome on continuous numeric scales.
Eventually, we obtained four candidate tools.
 
To enrich the list of potential candidates, we complemented the previous search by also looking for tools in GitHub. 
We used four search strings obtained by combining ``\textit{personality}'' with ``\textit{prediction}'' (258 entries), ``\textit{detection}'' (87), ``\textit{assessment}'' (47), and ``\textit{recognition}'' (38). 
Then, other than the filters applied earlier, we filtered out the repositories that (i) did not have an associated \texttt{README.md} file with installation and execution instructions and (ii) reported using personality models other than the FFM;
\revone{(iii) were supplementing material to references research papers (to exclude student projects).}

We identified only one candidate repository, which however was already included in the previous list.

Next, we review the four selected tools.
An overview is available in Table~\ref{tab:tools}, where we also report the dataset, techniques, and features used to develop the prediction model, as well as the instrument adopted to establish the ground truth.

Finally, while we cannot claim completeness---the systematic review of this research field is outside the scope of this study---this section provides nonetheless a valuable, up-to-date overview of the state of the art in the field of computational personality detection on continuous numeric scales.

\subsubsection{LIWC}
The Linguistic Inquiry and Word Count (pronounced \textit{luke})~\cite{pennebaker2015liwc} is a commercial, text-analysis program that counts words in psychologically meaningful, predetermined categories.
It adopts a top-down approach, therefore analyzing the text looking for the occurrence of predetermined linguistic cues associated with personality traits.
\liwc\ is arguably the most well-known resource in this category, often used as an external psycholinguistic database by other tools.
\fullcite{pennebaker1999linguistic} used \liwc\ to count the word categories of 2,479 essays (i.e., unedited pieces of text) written by volunteers who had also taken the BFI test as ground truth. 
In line with the lexical hypothesis, they found significant associations between the linguistic features of \liwc\ and the Big Five traits, thus providing evidence of existing connections between language use and personality~\cite{tausczik2010psychological}.

\textit{Output interpretation.} When analyzing a piece of text, \liwc\ returns word-category frequencies. We apply the formulae proposed by \fullcite{yarkoni2010liwc}, one for each of the big five traits, which leverage the quantified connections between personality and word use, and transform such frequencies into numerical trait scores in the range $1-5$.

\textit{Tool setup.} We used the standalone version of \liwc, which is also available from the web using the Receptivity\footnote{https://receptiviti.com} API.
Because the tool now supports both the 2007 and 2015 versions of the vocabulary, we opted for the first one, because it was the version used in \cite{yarkoni2010liwc} to derive the formulae. 
Nonetheless, despite a few differences among the categories defined, in our internal tests, the scores generated with the two dictionaries have very strong Pearson correlations (between .92 and .97).

\subsubsection{IBM Personality Insights}
It is a commercial tool that uses an unspecified machine-learning model with a bottom-up, open-vocabulary approach.
Earlier versions of the service (i.e., before December 2016) instead used the top-down, closed-vocabulary approach and relied on the \liwc\ dictionary. 
Models based on the open-vocabulary approach have been found to work well also in presence of small amounts of text such as tweets~\cite{arnoux2017baseline}. 
Also, as per IBM release note,\footnote{ \url{https://cloud.ibm.com/docs/personality-insights/science.html\#researchPrecise}} this version of \pin\ reportedly outperformed the previous \liwc-based model.
In November 2020, IBM announced\footnote{\url{https://cloud.ibm.com/docs/personality-insights?topic=personality-insights-release-notes}} that  the service had been deprecated and that it would be retired at the end of 2021.

\textit{Output interpretation.} After analyzing a piece of text, \pin\ returns a JSON response. We parse\revone{d} the JSON and retrieve five raw scores---one for each of the big five traits---defined in $[0,1]$, which we rescale\revone{d} in the range $1-5$. 

\textit{Tool setup.} Software Development Kits in multiple programming languages are available for the \pin\ service. In particular, we chose the Python API.

\subsubsection{Personality Recognizer} Top-down solutions make heavy use of external resources and test the correlations between those resources and personality traits.
The seminal work for top-down solutions is \pr,\footnote{\url{http://s3.amazonaws.com/mairesse/research/personality/recognizer.html}} a tool developed by \fullcite{mairesse2007pr} who conducted a series of experiments where multiple statistical models were benchmarked. 
With a supervised learning approach, they developed multiple prediction models using the same annotated dataset of essays employed for the development of \liwc. However, other than using \liwc\ features, they augmented the models with other dimensions from the Medical Research Council (MRC) psycholinguistic database~\cite{coltheart1981mrc}.

\textit{Output interpretation.} \pr\ issues trait predictions on a continuous, seven-point scale.
Therefore, to allow for comparison with other tools, we rescale\revone{d} the output in the range $1-5$.

\textit{Tool setup.} \pr\ requires the MRC database and the \liwc\ 2001 dictionary. In our experiment,  for consistency, we used the 2007 edition of \liwc. 
In addition, \pr\ supports different models for computing scores; we opted for the default option, i.e., Support Vector Machine (SVM) with Linear kernel (SMOreg). 
Finally, the tool can be optimized for the analysis of spoken or written language. Given the nature of our dataset, we chose the second option.

\subsubsection{TwitPersonality}
Another solution relying on the bottom-up approach is \tp.\footnote{\url{https://github.com/D2KLab/twitpersonality}} 
To develop the tool, \fullcite{carducci2018twitpers} used a supervised learning approach.
They first built a word-vector representation of Facebook posts (using the myPersonality dataset~\cite{stillwell2004mypersonality}) and then used it to train five SVM models, one for each trait. 
They also tested the models using a smaller corpus of tweets collected from 24 Twitter users, who also took the BFI test.
Albeit with some tinkering, \tp\ is the only tool among those benchmarked in this study, whose model can be retrained on new data.

\textit{Output interpretation.} \tp\ issues out-of-the-box trait predictions on a continuous, five-point scale. Therefore, no further transformations \revone{were} necessary.

\textit{Tool setup.} We used the default settings present in the source code. \tp\ can be used in two modes: \textit{user-wise}, i.e., the written `documents' of one author are aggregated and analyzed to extract the trait scores; \textit{post-wise}, i.e., the trait scores are inferred from the documents individually, and then the average scores are computed. We opted for the former mode, for the sake of consistency with the other tools, albeit our internal tests showed that the differences between the two modes, when present, are negligible.

\subsection{Gold Standard}
\label{sec:goldstandard}

In this section, we describe the process followed to build the gold standard and set the ground truth with a self-assessment questionnaire.
We retrieved the publicly available mailing lists archives of the projects belonging to the ASF,\footnote{\url{https://mail-archives.apache.org/mod_mbox}} as of Jan. 2018.
\revone{Our decision to investigate the ASF was motivated by the observation that all the projects within the ecosystem---albeit varying in size, scope, and technology stack---share the same code of conduct,\footnote{\url{www.apache.org/foundation/policies/conduct.html}} which enforces a shared set of guidelines that also regulate written interaction.}
\revone{Accordingly,} \revone{we focused on analyzing} the \texttt{dev} mailing lists because they are intended to host developers' discussions. 

We retrieved the list of all email addresses present in the archives and randomly selected 1,000 among those who had contributed at least 10,000 words in their emails, to ensure they had contributed enough text for the analysis.
We manually vetted the list to exclude the presence of emails automatically generated by bots such as the project's version control system or the mailing-list software.
Finally, we sent invitations by email to the selected developers to take a personality test. 

To collect the responses, we developed an electronic version of the 20-item Mini-IPIP~\cite{donnellan2006miniipip} questionnaire---the shortest, valid personality instruments available---in an attempt to increase the notoriously low response rate of surveys in the \SE\ domain~\cite{smith2013sesurveys}.
The form collected the responses and, following the specifications provided in~\cite{donnellan2006miniipip},  transformed them into trait scores in the range $1-5$.
In addition to the test questions, we inserted a couple of attention items and also measured the time taken to complete the tests.
No monetary incentives were given to the test participants.

\revone{In the 2010s, personal data belonging to millions of Facebook users was collected without their consent by the consulting firm Cambridge Analytica and used during the 2016 US presidential campaigns for \textit{psychological targeting}, i.e., the extraction of psychological profiles from social-media digital footprints to influence the attitudes, emotions, and behaviors of large groups of people.
This scandal increased the public attention to the privacy risks of personal data misuse, creating still persistent social stigma attached to personality-related research and concerns deriving from participating in related studies~\cite{matz2020privacy}.
Therefore, g}iven the sensitive nature of the data collected in the study, both the invitation emails and the website contained a detailed description of the goal of the study and its academic-only interest.
We ensured the developers taking the test that the results would only be presented in aggregate and that no resource would be shared, which could allow third parties to match the test results to their identities.
For further details on anonymity and data protection measures adopted during data collection, please refer to the replication package in Appendix~\ref{appendix:rep-package}. 

\begin{figure}[t]
\centerline{
\includegraphics[width=0.7\textwidth]{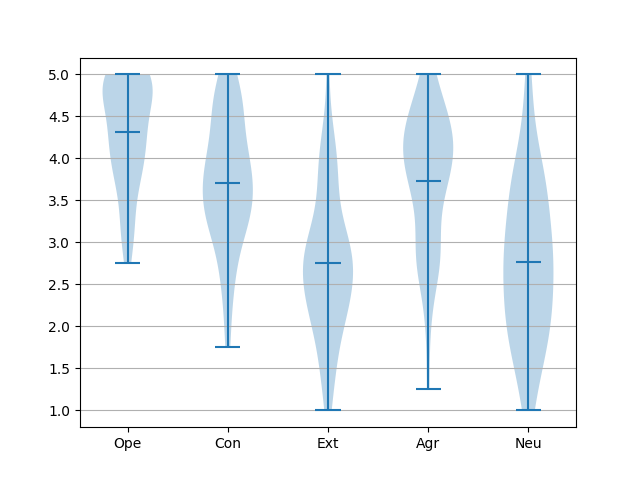}}
\vspace{-5mm}
\caption{Distribution of the self-ratings from the personality questionnaire.}
\label{fig:violins-gs}
\end{figure}

We received 61 responses (6\% response rate), of which 50 were deemed valid. 
Seven responses were excluded because the respondents failed the attention checks and four because of the short time spent in taking the test (less than two minutes over an average of nearly eight).
\revone{The survey respondents belong to 34 different ASF projects,\footnote{The complete list of projects and respondents is available at \url{https://doi.org/10.6084/m9.figshare.15066564}.} including \texttt{lucene} (4 developers), \texttt{maven} (4), \texttt{couchdb} (3), \texttt{log4net} (3), \texttt{kafka} (2), \texttt{cassandra} (2), and \texttt{openmeetings} (2).}
From Fig.~\ref{fig:violins-gs}, we observe that the participants tend to be \textit{open} (mean 4.33, SD  .63), exhibit average levels of \textit{conscientiousness} (3.70, .75) and \textit{agreeableness} (3.73, .80), and are neither very \textit{extroverted} (2.73, .84) or \textit{neurotic} (2.77, .92).

\subsection{Experimental Dataset}
\label{sec:dataset}

We matched the self-assessed personality profiles with a corpus of all the emails written by the developers who took the test.
As a result, we were able to run the selected tools to infer the personality scores from the email corpus and compare the predictions against the self-ratings from the questionnaire (gold standard).

To build the corpus, we  aggregated all the email bodies sent by each subject and applied a series of filters to clean the data.
In particular, we first used the \texttt{email-reply-parser}\footnote{\url{https://pypi.org/project/email-reply-parser}} library to ensure that only the text written by the email author was retrained while discarding the signature and any text coming from replying and forwarding.
Then, we removed the lines of code using the R package \texttt{NLoN}, developed by \fullcite{mantyla2018nlon}. 
\texttt{NLoN}  was trained and tested on various email corpora---including one from Apache developers---with good results.
In addition, we used \texttt{polyglot}\footnote{\url{https://pypi.org/project/polyglot}} to remove any non-English words.
Finally, we lower-cased the text and removed all the stopwords using \texttt{NLTK}.\footnote{\url{www.nltk.org}}

Overall, the 50 subjects contributed 1,543 emails, with an average of \revone{30.86} per developer (\revone{min 15, max 55, median 33,} SD 8.45).
After collating the email bodies and performing the data cleaning, we \revone{found} that each developer has contributed on average 1,111\revone{.04} words (\revone{min 747, max 1764, median 1098}, SD 180.14).
We notice that these dimensions are in line with those of other datasets used in training and testing the selected tools. 
For example, \pin\ recommends providing at least between 600 and 1,200 words to enable the analysis,\footnote{\url{https://cloud.ibm.com/docs/personality-insights?topic=personality-insights-input}} albeit it is  not disclosed what kind of preprocessing is applied. 
The dataset used by \liwc\ and \pr\ contains 2,479 essays with an average length of 652 words.
\tp\ employed a dataset of tweets, most of which are typically a hundred characters long.\footnote{\url{https://blog.twitter.com/official/en_us/topics/product/2017/tweetingmadeeasier.html}}

\subsection{Evaluation Metrics}
\label{sec:metrics}
Research on the Big Five model has consistently considered personality detection as a set of five separate trait-prediction tasks.
However, the formulation of the prediction tasks can differ drastically~\cite{oberlander2006tasks}:
they can be approached as binary classification tasks, using the mean or median as thresholds to discretize the numerical scores;
alternatively, after removing the observations in the middle, they approach the binary classification as limited to the upper and lower groups, albeit this is less than ideal with bell-shaped distributions.

The choice of the metrics for evaluating performance also varies. In the case of personality predictions on a continuous scale, following~\cite{schwartz2013openvocabulary}, in our analysis, we include\revone{d} the following performance metrics: \textit{Pearson} product-moment correlation ($r$) and \textit{Spearman} rank correlation ($\rho$),  measured between the predicted trait scores (by each tool) and the actual scores (from the gold standard, self-assessment questionnaire); \textit{Mean Absolute Error} (MAE),\footnote{$MAE_t=\frac{\sum_{i=i}^n |y_i - \widehat{y_i}|}{n}$ where $t$ is the trait, and $y_i$ and $\widehat{y_i}$ are the ground truth and predicted scores for subject $i=1..n$.} the average of the absolute value of the difference between the actual and predicted scores; \textit{Root Mean Squared Error} (RMSE),\footnote{$RMSE_t=\sqrt{\frac{\sum_{i=i}^n(y_i - \widehat{y_i})^2}{n}}$ where $t$ is the trait, and $y_i$ and $\widehat{y_i}$ are the ground truth and predicted scores for subject $i=1..n$.} the standard deviation of the residuals, i.e., the prediction errors.

\begin{table}[t]
\tiny
\revonetable
\caption{Reference values in terms of Pearson $r$ and Spearman $\rho$ correlations from prior work that employed text-based datasets (larger is better).}
\label{tab:correlations_comparisons}
\begin{tabular}{lcccccccccc}
\hline
\multicolumn{1}{c}{\multirow{2}{*}{\textbf{\begin{tabular}[c]{@{}c@{}}Study\\ (best results)\end{tabular}}}} & \multirow{2}{*}{\textbf{Approach}} & \multirow{2}{*}{\textbf{Technique}}                                      & \multirow{2}{*}{\textbf{Subjects}} & \multirow{2}{*}{\textbf{Dataset}} & \multirow{2}{*}{\textbf{\begin{tabular}[c]{@{}c@{}}Validation\\ (ground truth)\end{tabular}}} & \multicolumn{5}{c}{\textbf{Pearson $r$}}                           \\ \cline{7-11} 
\multicolumn{1}{c}{}                                                                                         &                                    &                                                                          &                                    &                                   &                                                                                               & \textbf{O} & \textbf{C}     & \textbf{E} & \textbf{A} & \textbf{N} \\ \hline
\fullcite{lynn2020baseline}                                                                                  & BU                                 & \begin{tabular}[c]{@{}c@{}}Message-level \\ attention model\end{tabular} & 68,687                             & myPersonality                     & \begin{tabular}[c]{@{}c@{}}myPersonality\\ (IPIP)\end{tabular}                                & 0.66       & 0.54           & 0.58       & 0.56       & 0.56       \\
\fullcite{laleh2017baseline}                                                                                 & TD                                 & \begin{tabular}[c]{@{}c@{}}Regression\\ analysis\end{tabular}            & 92,225                             & myPersonality                     & \begin{tabular}[c]{@{}c@{}}myPersonality\\ (IPIP)\end{tabular}                                & 0.38       & 0.29           & 0.34       & 0.33       & 0.27       \\
\fullcite{arnoux2017baseline}                                                                                & TD                                 & \begin{tabular}[c]{@{}c@{}}Regression\\ analysis\end{tabular}            & 1,323                              & Private                           & IPIP                                                                                          & 0.29       & 0.33           & 0.27       & 0.42       & 0.37       \\
\fullcite{kosinski2013baseline}                                                                              & TD                                 & \begin{tabular}[c]{@{}c@{}}Regression\\ analysis\end{tabular}            & 54,373                             & myPersonality                     & \begin{tabular}[c]{@{}c@{}}myPersonality\\ (IPIP)\end{tabular}                                & 0.43       & 0.29           & 0.40       & 0.30       & 0.30       \\
\fullcite{tomlinson2013baseline}                                                                             & TD                                 & \begin{tabular}[c]{@{}c@{}}Regression\\ analysis\end{tabular}            & 250                                & myPersonality                     & \begin{tabular}[c]{@{}c@{}}myPersonality\\ (IPIP)\end{tabular}                                & -          & 0.27 & -          & -          & -          \\ \hline
\multicolumn{1}{c}{\multirow{2}{*}{\textbf{\begin{tabular}[c]{@{}c@{}}Study\\ (best results)\end{tabular}}}} & \multirow{2}{*}{\textbf{Approach}} & \multirow{2}{*}{\textbf{Technique}}                                      & \multirow{2}{*}{\textbf{Subjects}} & \multirow{2}{*}{\textbf{Dataset}} & \multirow{2}{*}{\textbf{\begin{tabular}[c]{@{}c@{}}Validation\\ (ground truth)\end{tabular}}} & \multicolumn{5}{c}{\textbf{Spearman $\rho$}}                       \\ \cline{7-11} 
\multicolumn{1}{c}{}                                                                                         &                                    &                                                                          &                                    &                                   &                                                                                               & \textbf{O} & \textbf{C}     & \textbf{E} & \textbf{A} & \textbf{N} \\ \hline
\fullcite{hall2017socialmedia}                                                                               & TD                                 & \begin{tabular}[c]{@{}c@{}}Regression\\ analysis\end{tabular}            & 509                                & Private                           & \begin{tabular}[c]{@{}c@{}}Unspecified Big 5\\ questionnaire\end{tabular}                     & 0.62       & 0.68           & 0.70       & 0.68       & 0.61       \\ \hline
\multicolumn{11}{l}{\textit{TD stands for Top-Down (closed vocabulary), BU for Bottom-Up (open vocabulary).}} 
\vspace{-5mm}
\end{tabular}
\end{table}

Pearson correlation evaluates the linear relationship between two continuous variables. Coefficients can range from -1 (perfect negative) to +1 (perfect positive), with values close to 0 indicating the lack of correlation. 
Assessing the predictive performance in automatic personality detection means estimating the convergent validity, i.e., the degree to which the measures of the same construct correlate with each other. 
One problem with using correlations is how to interpret the results.  
A common interpretation of the observed correlation magnitude is: $.00-.10$ negligible, $.10-.39$ weak, $.39-.69$ moderate, $.69-.89$ strong, $.89-1.0$ very strong~\cite{schober2018corr-cutoffs}. 
However, as observed by the authors, these cutoff points are arbitrary and should be used judiciously. 
In particular, values in the middle are disputable and their interpretation as weak, moderate, or strong varies with the applied rule of thumb.
Achieving correlations of $r>.30$ in psychology studies is challenging---even the simple axiom according to which people's past behavior is predictive of future actions has been found to produce a correlation coefficient of $r\approx.39$ \cite{meyer2001corr}.
Rather than relying on the conventional cut-off points used for interpreting correlation coefficients in other fields, \fullcite{meyer2001corr} and \fullcite{roberts2007corr} have argued that research investigating psychological constructs should use baselines in the order of magnitude of correlations independently measured in related work. 
In other words, they have called for adjusting the norms that researchers hold for what the strength of relationships is in psychology and related fields.
For example, \pin\ reportedly achieved for English an average Pearson correlation coefficient  $r\approx.33$ in an internal assessment study.
Some studies on psychological and behavioral constructs (e.g., \cite{roberts2007corr}) have reported Pearson correlations with a small to medium magnitude in the range $.10-.40$.
In a survey of over 200 papers on personality published since 2017, \fullcite{novikov2021survey} found that the reported Pearson correlation coefficients between predicted and self-reported personality traits are upper limited by values near $.50$.
One exception is represented by the work of~\fullcite{lynn2020baseline}, which reports a score exceeding 0.60 in the case of \textit{openness}.
Table~\ref{tab:correlations_comparisons} lists some of the Pearson correlation coefficients reported in recent prior work.
We will use these values as baselines to assess the performance of the personality prediction tools involved in our study.
\revone{We notice that most of the studies that reported Person correlation metrics adopted a top-down approach, using regression analyses for analyzing data from the myPersonality dataset.}

Spearman correlation coefficient also varies between -1 and +1.
Unlike Pearson $r$, however, Spearman $\rho$ is a non-parametric measure that does not make any assumption regarding the normality, linearity, and homoscedasticity of distributions.
The Spearman coefficient is based on the ranked values for each variable rather than the raw data.
\fullcite{fowler1987spearman} found Spearman rank correlations to be more robust and outperforming Pearson $r$ in cases of non-normal distributions.
Only a few studies (e.g.,~\cite{hall2017socialmedia}) have assessed prediction performance using Spearman~$\rho$, which appears to be used more frequently with multimedia datasets.
A complete analysis of correlation coefficients reported in prior work is out of the scope of this work.
For more, please refer to the meta-analyses performed by \fullcite{azucar2018meta} and \fullcite{marengo2020meta}.

The MAE and RMSE measures are also very popular.
\fullcite{novikov2021survey} found \revone{that 65 out of the 218 studies analyzed} report personality prediction performance using either measure. 
Nonetheless, these metrics are not without problems.
Their interpretation is largely dependent on the scale of the data and their estimates tend to be over-optimistic since self-ratings in gold standards tend to be normally distributed, with most observations close to the mean \cite{sumner2012rmse-mae}.
Given their popularity, we include these metrics for the sake of comparison with prior work. Table~\ref{tab:error_comparisons} provides an overview of the best MAE and RMSE results reported in recent and relevant studies, thus providing us with a performance baseline.
\revone{We notice that the top-down and bottom-up approaches are almost equally distributed and that these studies mostly relied on regression analyses. However, while the studies reporting RMSE typically used myPersonality as the data source and ground truth, the others that relied on MAE built private datasets, using the IPIP or BFI questionnaires for validation.}

\begin{table}[t]
\tiny
\revonecaption
\revonetable
\caption{Reference values in terms of Mean Absolute Error (MAE) and Root Mean Square Error (RMSE) from prior work that employed text-based datasets (smaller is better).}  
\label{tab:error_comparisons}
\begin{tabular}{lcccccccccc}
\hline
\multicolumn{1}{c}{\multirow{2}{*}{\textbf{\begin{tabular}[c]{@{}c@{}}Study\\ (best results)\end{tabular}}}} & \multirow{2}{*}{\textbf{Approach}} & \multirow{2}{*}{\textbf{Technique}}                            & \multirow{2}{*}{\textbf{Subjects}} & \multirow{2}{*}{\textbf{Dataset}}                                & \multirow{2}{*}{\textbf{\begin{tabular}[c]{@{}c@{}}Validation\\ (ground truth)\end{tabular}}} & \multicolumn{5}{c}{\textbf{MAE}}                                                                                                          \\ \cline{7-11} 
\multicolumn{1}{c}{}                                                                                         &                                    &                                                                &                                    &                                                                  &                                                                                               & \textbf{O}                & \textbf{C}                & \textbf{E}                & \textbf{A}                & \textbf{N}                \\ \hline
\pin                                                                                                         & BU                                 & Unspecified                                                    & -                                  & Private                                                          & \begin{tabular}[c]{@{}c@{}}Unspecified Big 5\\ questionnaire\end{tabular}                     & \multicolumn{5}{c}{Avg. $\approx0.120$}                                                                                                   \\
\fullcite{goldbeck2011chi}                                                                                   & TD                                 & \begin{tabular}[c]{@{}c@{}}Regression \\ analysis\end{tabular} & 167                                & Private                                                          & BFI                                                                                           & 0.099                     & 0.104                     & 0.124                     & 0.109                     & 0.117                     \\
\fullcite{goldbeck2011twitter}                                                                               & TD                                 & \begin{tabular}[c]{@{}c@{}}Regression \\ analysis\end{tabular} & 50                                 & Private                                                          & BFI                                                                                           & 0.130                     & 0.146                     & 0.160                     & 0.182                     & 0.119                     \\
\fullcite{arnoux2017baseline}                                                                                & BU                                 & \begin{tabular}[c]{@{}c@{}}Regression\\ analysis\end{tabular}     & 1,323                              & Private                                                          & IPIP                                                                                          & \multicolumn{5}{c}{Avg. $\approx0.120$}                                                                                                   \\ \hline
\multicolumn{1}{c}{\multirow{2}{*}{\textbf{\begin{tabular}[c]{@{}c@{}}Study\\ (best results)\end{tabular}}}} & \multirow{2}{*}{\textbf{Approach}} & \multirow{2}{*}{\textbf{Technique}}                            & \multirow{2}{*}{\textbf{Subjects}} & \multirow{2}{*}{\textbf{Dataset}}                                & \multirow{2}{*}{\textbf{\begin{tabular}[c]{@{}c@{}}Validation\\ (ground truth)\end{tabular}}} & \multicolumn{5}{c}{\textbf{RMSE}}                                                                                                         \\ \cline{7-11} 
\multicolumn{1}{c}{}                                                                                         &                                    &                                                                &                                    &                                                                  &                                                                                               & \textbf{O}                & \textbf{C}                & \textbf{E}                & \textbf{A}                & \textbf{N}                \\ \hline
\begin{tabular}[c]{@{}l@{}}\fullcite{carducci2018twitpers}\\ (\tp)\end{tabular}                              & BU                                 & SVM                                                            & 24/250                             & \begin{tabular}[c]{@{}c@{}}Private/\\ myPersonality\end{tabular} & \begin{tabular}[c]{@{}c@{}}myPersonality\\ (IPIP)\end{tabular}                                 & 0.332                     & 0.530                     & 0.708                     & 0.448                     & 0.557                     \\
\fullcite{quercia2011twitter}                                                                                & TD                                 & \begin{tabular}[c]{@{}c@{}}Regression \\ analysis\end{tabular} & 335                                & Private                                                          & \begin{tabular}[c]{@{}c@{}}myPersonality\\ (IPIP)\end{tabular}                                 & 0.690                     & 0.760                     & 0.880                     & 0.790                     & 0.850                      \\
\fullcite{laleh2017baseline}                                                                                 & TD                                 & \begin{tabular}[c]{@{}c@{}}Regression\\ analysis\end{tabular}  & 92,225                             & myPersonality                                                    & \begin{tabular}[c]{@{}c@{}}myPersonality\\ (IPIP)\end{tabular}                                 & \multicolumn{1}{l}{0.155} & \multicolumn{1}{l}{0.173} & \multicolumn{1}{l}{0.195} & \multicolumn{1}{l}{0.173} & \multicolumn{1}{l}{0.195} \\ \hline
\multicolumn{11}{l}{\textit{TD stands for Top-Down (closed vocabulary), BU for Bottom-Up (open vocabulary).}}  
\end{tabular}
\end{table}

\subsection{Results}
\label{sec:ph1-results}

\begin{figure}
\centering
    \subfloat[LIWC\label{fig:violins_liwc}]{%
       \includegraphics[width=0.5\textwidth]{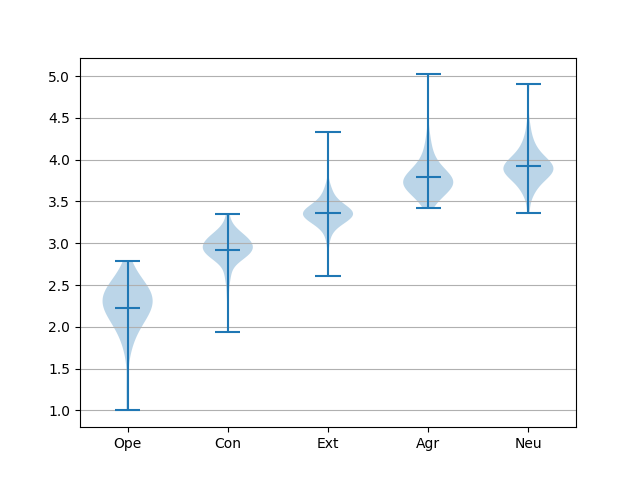}
     }
    \subfloat[Personality Insights\label{fig:violins_pi}]{%
       \includegraphics[width=0.5\textwidth]{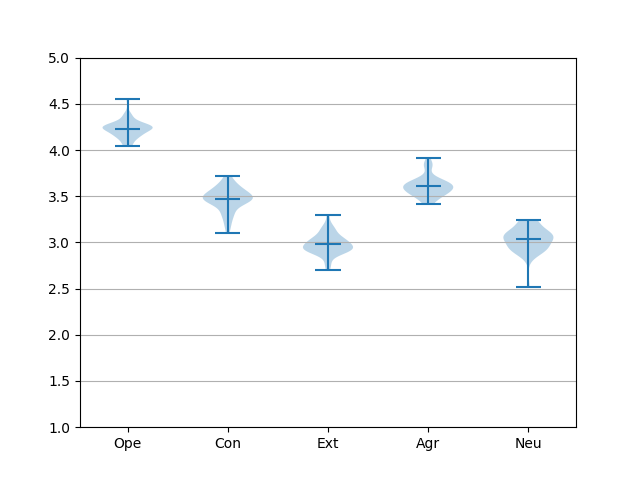}
     }
     \hfill\vspace{-5mm}
     \subfloat[Personality Recognizer\label{fig:violins_pr}]{%
       \includegraphics[width=0.5\textwidth]{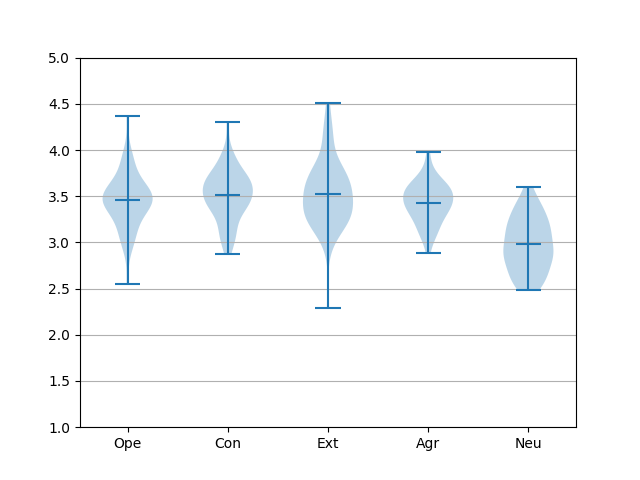}
     }
     \subfloat[TwitPersonality\label{fig:violins_tp}]{%
       \includegraphics[width=0.5\textwidth]{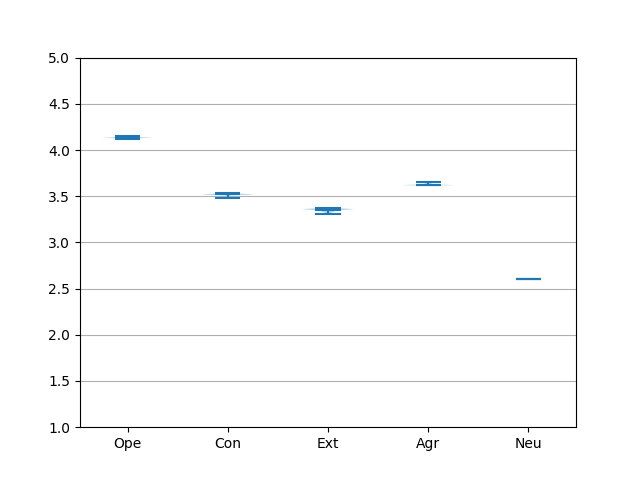}
     }
\caption{Violin plots of the trait predictions for each of the selected tools.}
\label{fig:violins_tools}
\end{figure}

In Figures~\ref{fig:violins_tools}a-d we report the violin plots of the trait score distributions inferred by the tools.
As compared to the distributions of the gold standard scores reported earlier in Fig.~\ref{fig:violins-gs}, we notice that the tool predictions are far less spread out.
This can be also observed from the small standard deviations reported along with other descriptive statistics in Appendix~\ref{appendix:additional-material-ph1} (see Table~\ref{tab:desc-stats}).
In particular, we observe that \tp\ issues predictions for each score that are clustered around the mean, and even constant in the case of \textit{neuroticism}.
The Q-Q plots in Figures~\ref{fig:qqplot-gs}-\ref{fig:qqplot-tp} (see Appendix~\ref{appendix:additional-material-ph1}) show that while linear relationships exist in the distributions of the self-ratings as well as the tool predictions, they are not normal, thus violating one of the assumptions to apply Pearson $r$ correlation.

\subsubsection*{Agreement Analysis}
\label{sec:agreement-analysis}
We assess\revone{ed} the level of agreement by comparing the personality scores predicted by the tools, respectively, against the self-reported ratings and between each other.
The Pearson and Spearman correlation coefficients are reported in Table~\ref{tab:correlations_results}. 
We point out that the correlation coefficients could not be calculated for \textit{neuroticism} in the case of \tp\ because the tool issues a constant value for all the subjects.

Overall, we notice that no pair does consistently better than any other in terms of either correlation coefficient.
Regarding Pearson correlation, the highest $r$ coefficient is computed between \liwc\ and \tp\ for the \textit{extraversion} trait ($r=.34$).
However, most coefficients are approximately equal to .10 or smaller, and several are negative.
Consistently, we observe that the largest $r$ coefficients between all pairs and across all traits in Table~\ref{tab:correlations_results} ($r\approx.05-.34$) are smaller than those found in prior work and reported earlier in Table~\ref{tab:correlations_comparisons} ($r\approx.40-.60$).
In terms of Spearman correlation, the same observations still hold, with $\rho$ coefficients in the range $.10 - .34$, therefore smaller than those reported in prior work ($\rho\approx.60-.70$).

Concerning the correlations between the gold standard and the tools, we cannot identify any pattern. 
Instead, regarding the correlations between tools, we notice that most of the largest $r$ and $\rho$ coefficients are obtained for pairs including \liwc. 
In the case of \liwc\ and \pr, this result is not surprising since the latter uses the \liwc\ dictionary to build the prediction model and, therefore, the two tools share some linguistic features.

\begin{table}[t]
\small
\caption{Pearson $r$ and Spearman $\rho$ pairwise correlations for each of the five traits between the gold standard (GS) and each tool, and between tools. The best result (highest positive correlation) for each trait is reported in \textbf{bold}.}
\label{tab:correlations_results}
\begin{tabular}{lccccc|ccccc}
\hline
\multirow{2}{*}{\textbf{Pair}}               & \multicolumn{5}{c|}{\textbf{Pearson $r$}}                                          & \multicolumn{5}{c}{\textbf{Spearman $\rho$}}                                       \\ \cline{2-11} 
                                         & \textbf{O}     & \textbf{C}     & \textbf{E}     & \textbf{A}     & \textbf{N}     & \textbf{O}     & \textbf{C}     & \textbf{E}     & \textbf{A}     & \textbf{N}     \\ \hline
GS -- \liwc                              & -0.063         & -0.057         & -0.325         & -0.151         & -0.090         & 0.030          & -0.056         & -0.151         & \textbf{0.134} & -0.048         \\
GS -- \pin                               & -0.050         & \textbf{0.016} & -0.156         & -0.001         & -0.073         & -0.048         & \textbf{0.140} & -0.200         & 0.009          & -0.107         \\
GS -- \textsc{Pers.Rec.}                 & -0.036         & -0.028         & \textbf{0.063} & -0.120         & \textbf{0.142} & \textbf{0.033} & -0.125         & \textbf{0.034} & 0.041          & \textbf{0.083} \\
GS -- \textsc{TwitPers.}                 & \textbf{0.016} & -0.148         & -0.114         & \textbf{0.150} & -              & -0.011         & -0.148         & -0.137         & 0.131          & -              \\ \hline
\liwc\ -- \pin\                          & -0.169         & \textbf{0.046} & 0.112          & -0.144         & \textbf{0.173} & -0.134         & \textbf{0.003} & 0.213          & -0.148         & \textbf{0.132} \\
\liwc\ -- \textsc{Pers.Rec.}             & \textbf{0.212} & -0.185         & -0.025         & \textbf{0.310} & -0.175         & \textbf{0.102} & -0.073         & 0.035          & \textbf{0.100} & -0.035         \\
\liwc\ -- \textsc{TwitPers.}             & 0.055          & -0.006         & \textbf{0.343} & -0.023         & -              & 0.052          & -0.037         & \textbf{0.217} & -0.100         & -              \\
\pin\ -- \textsc{Pers.Rec.}              & -0.037         & -0.072         & 0.191          & -0.064         & -0.003         & -0.037         & -0.099         & 0.159          & -0.092         & -0.015         \\
\pin\ -- \textsc{TwitPers.}              & -0.195         & -0.224         & -0.050         & -0.104         & -              & -0.183         & -0.232         & 0.000          & -0.198         & -              \\
\textsc{Pers.Rec.} -- \textsc{TwitPers.} & -0.073         & -0.043         & -0.175         & 0.090          & -              & -0.077         & -0.027         & -0.082         & 0.099          & -              \\ \hline
\end{tabular}
\end{table}

\subsubsection*{Accuracy analysis}
\label{sec:accuracy-analysis}

To complete the performance assessment, we complement\revone{ed} the agreement analysis by making sense of the prediction accuracy.  
Table~\ref{tab:error_results} reports the MAE and RMSE metrics for each tool.
First, we observe that there is no best tool in absolute, which outperforms all the others.
\liwc\ does better for the \textit{extraversion} and \textit{agreeableness} traits; \tp\ performs better at estimating \textit{conscientiousness} and \textit{neuroticism}.
However, a known limitation of MAE and RMSE is failing to capture the performance accurately when distributions containing observations that are mostly clustered around the mean, as in the case of \tp.
Nevertheless, the best results in this study (MAE~$\approx.55-.77$, RMSE $\approx .42-.90$) are considerably larger (worse) than the values found in prior work and reported earlier in Table~\ref{tab:error_comparisons} (MAE $\approx.10-.18$, RMSE $\approx .16-.88$).

\begin{table}[t]
\caption{Mean Absolute Error (MAE) and Root Mean Square Error (RMSE). The best (smallest) result for each trait is reported in \textbf{bold}.}
\label{tab:error_results}
\small
\begin{tabular}{lccccc|ccccc}
\hline
\multirow{2}{*}{\textbf{Tools}} & \multicolumn{5}{c|}{\textbf{MAE}}                              & \multicolumn{5}{c}{\textbf{RMSE}}                              \\ \cline{2-11} 
                                & \textbf{O} & \textbf{C} & \textbf{E} & \textbf{A} & \textbf{N} & \textbf{O} & \textbf{C} & \textbf{E} & \textbf{A} & \textbf{N} \\ \hline
\liwc                           & 2.643      & 1.268      & \textbf{0.709}      & \textbf{0.668}      & 1.200      & 7.441      & 2.146      & \textbf{0.759}      & \textbf{0.653}      & 2.059      \\ 
\pin                            & \textbf{0.554}      & 0.632      & 0.720      & 0.678      & 0.798      & \textbf{0.424}      & 0.632      & 0.800      & 0.674      & 0.952      \\ 
\textsc{Pers. Rec.}             & 0.955      & 0.628      & 1.011       & 0.770      & \textbf{0.767}      & 1.233      & 0.650      & 1.376      & 0.795      & 0.925      \\ 
\textsc{TwitPers.}              & 0.566      & \textbf{0.611}      & 0.882      & 0.676      & 0.769      & 0.431      & \textbf{0.606}      & 1.065      & 0.663      & \textbf{0.889}      \\ \hline
\end{tabular}
\end{table}

\subsection{Threats to Validity}
\label{sec:threats-phase1}


The results of the analyses should be interpreted in light of the following limitations.

First, while using individual self-ratings as gold standards to set ground truth is the norm, psychology research considers the definition of a ‘true’ personality profile out of reach~\cite{wright2014ph1limitations}. 
Indeed, personality is an elusive concept whose assessment makes it a complex activity for any rater, whether self, external observer\revone{,} or computer.
Any form of personality rating is a proxy measure and, as such, it comes with its limitations.
Despite their validity, self-assessment questionnaires are subjective and biased towards social desirability~\cite{boyle2009ph1limitations}; external judgments are limited by raters' idiosyncrasies and machine learning algorithms, while free of human prejudice, reflect biases present in the data~\cite{tay2020validity}.
Also, albeit highly correlated, there are differences between personality constructs based on self-ratings and external observers' judgments~\cite{mount1994ph1limitations}.

In line with recent meta-reviews of personality prediction model performance~\cite{azucar2018meta,marengo2020meta}, we used correlation coefficients (both Pearson and Spearman) as reference metrics.
We compared studies that were trained and tested on different datasets (e.g., essays, social media posts). 
The heterogeneity of the data sources might have influenced the comparability of the results.
However, this was intentional to assess 
how personality detection tools can be used across domains as off-the-shelf solutions, and specifically for \SE\ research.

Another potential issue is related to the use of English as lingua franca in emails, i.e., some developers did not communicate using their native language.
A limited vocabulary may have arguably prevented some lexical cues from emerging from the text, as argued in the lexical hypothesis.

Finally, we acknowledge the relatively low number of subjects (50 developers) and documents (631 emails) in our experimental sample.
However, it is not uncommon to find previous studies on personality detection using samples of a similar size.
For instance, \fullcite{carducci2018twitpers} tested their personality detection tool using a corpus of Twitter posts from 24 volunteers.
Similarly, \fullcite{arnoux2017baseline} report the Big Five traits extracted from the social media posts of 55 volunteers.


\begin{tcolorbox}[standard jigsaw,
	title=Phase 1 -- Summary, opacityback=0]
    When  off-the-shelf personality detection tools  are applied out of domain, their performance is far from acceptable.
    Indeed, the tools neither agree with self-reported ratings nor with each other when used for \SE\ research. 
    In addition, their prediction accuracy is considerably worse as compared to the results from prior work on automatic personality detection from text.
\end{tcolorbox}


\section{Phase 2 -- Implications on Earlier Studies}
\label{sec:replicated-studies}
The results of Phase 1 suggest that, due to the limited level of agreement and accuracy of tool predictions, the choice of a personality detection tool \textit{might} affect the validity of previous results.
Indeed, these disagreements do not necessarily imply that conclusions based on the application of these tools in the software engineering domain are affected by the choice of one specific tool over the others.

Accordingly, to answer RQ2, in this section we investigate whether the choice of a specific personality tool introduces threats to conclusion validity by replicating two previous studies in \SE, which relied on computational personality detection. 
Because our goal is to assess whether the effects reported in previous studies still hold when a different personality detection tool is used, here we perform two \textit{exact dependent replications}~\cite{shull2008replications} in which we keep the same experimental setup of the original studies while changing only the tool.

\subsection{Replicated Studies}
\label{sec:replications}

We choose to replicate two previous studies, respectively by \fullcite{iyer2019github} and \fullcite{calefato2019apache}, which have analyzed the personalities of open-source software developers. 
Both studies provide a replication package. 
Therefore, we can apply the chosen methodology and perform an exact replication of the studies with the only deviation from the original work being the different personality detection tool used. 
Since both original studies used \pin\, here we choose to replace it with \liwc\ because it is the most used psychometric tool, adopted also in previous studies on personality in \SE\ (e.g., \cite{rigby2007liwc-asf,bazelli2013liwc-so,rastogi2016liwc-gh}).

\subsubsection{\citeauthor{iyer2019github}~(\citeyear{iyer2019github})}

The first study that we replicate is by \fullcite{iyer2019github}. The authors applied \pin\ to examine the influence of personality traits of developers on the pull request evaluation process in GitHub.
They first extracted the Big Five personality traits of 16,935 developers from trace data on GitHub, such as commit messages, issue and pull request comments. Then, they assessed their relative importance in the pull request evaluation process as compared to other non-personality factors from past research. 
Overall, they evaluated 501,327 pull requests from 1,860 projects and found that the effect of personality traits is significant and comparable to technical factors (e.g., number of files changed, presence of tests), albeit social factors (e.g., prior interaction, following each other) are more influential on the likelihood of pull request acceptance.
In particular, they found that pull requests authored by developers (requesters) who are more \textit{open} and \textit{conscientious}, but less \textit{extroverted}, have a higher chance of acceptance. 
Furthermore, pull requests that are closed by developers (closers) who are more \textit{conscientious}, \textit{extroverted}, and \textit{neurotic}, have a higher likelihood of acceptance. 
Additionally, the larger the difference in personality traits between the requester and the closer, the more positive effect it has on pull request approval.

For the re-execution, we adapted the original scripts provided in the replication package. 
However, since the package did not include the collection of traced data used for the analysis, we followed the description of the data collection process in the paper to recreate the dataset ourselves. 
This lead to minor differences in the number of projects found (1,853) as compared to those reported in the original study (1,860). 
Finally, we applied \liwc\ to the reconstructed dataset to infer the developers' personality scores.

\subsubsection{\citeauthor{calefato2019apache}~(\citeyear{calefato2019apache})}

The second study that we replicate is by \fullcite{calefato2019apache}. 
The authors applied \pin\ to perform an analysis at the ecosystem-level of code commits and email messages contributed by 211 developers working on ASF projects. 
They found that there are three common types of personality profiles among Apache developers, characterized in particular by their level of \textit{agreeableness} and \textit{neuroticism}. 
They also found that developers with higher levels of \textit{openness} are more likely to become contributors to ASF projects.
In addition, they confirmed that developers' personality is stable over time and that the five traits do not vary significantly with their role, membership, and extent of contribution to the projects. 

For the replication, we adapt the same scripts and use the same experimental dataset from the original study on which we apply \liwc.

\subsection{Replication Results}
To distinguish from those of this work, the research questions of the original studies are formatted in italic and lower-cased (e.g., \textit{rq1}, \textit{rq2}, \ldots).
\label{sec:ph2-results}

\subsubsection{Replication of \citeauthor{iyer2019github}~(\citeyear{iyer2019github})} 

In this section, we answer the four research questions \textit{rq0-3} of the original study by \fullcite{iyer2019github} and recreate the same tables for comparison.

\textit{rq0---Replication of base model with the reconstructed dataset.}

Given the slight differences between the original and reconstructed datasets, we replicate \citeauthor{iyer2019github}'s findings of the baseline model---including only technical and social factors---on a more recently mined dataset to determine if the results still hold.
Creating a baseline model is useful in comparing other personality models with social, technical, and personality factors.
In addition, GitHub has tremendous yearly growth rates, and over 60M new repositories have been created since the original study was carried out.\footnote{\url{https://octoverse.github.com}, accessed in March 2021.} 
As such, the replication of the baseline model provides insights on the generalizability of \citeauthor{iyer2019github}'s results to a dataset extracted at a different point in time.

We use the same modeling technique as in the original study---a mixed-effects logistic regression---on the reconstructed dataset to assess the effects of social and technical factors on pull request acceptance. 
Table~\ref{tab:rep1_rq0_dataset} provides a comparison between our results and the original ones by \citeauthor{iyer2019github} in terms of odds ratios.
All factors have similar overall influences and even the model fit, both marginal ($R^2m$) and conditional ($R^2c$), is similar.
The only exception is the project age factor, for which we \revone{found} a significant effect.
We speculate that this is due to the extraction of projects that have been active for a longer time in the reconstructed dataset. 

Overall, although there are small fluctuations in the odds ratio of all the features, the original results still hold. 
This increases the confidence in the goodness of the reconstructed dataset and that any statistically significant, different results in the replication of the other research questions are due to the different personality detection tool used rather than differences in the data.

\begin{table}[t]
\caption{Odds ratio of the basic models without personality traits from the original study by \fullcite{iyer2019github} and the replication. Significant values are shown in \textbf{bold} (sig: *$p < .05$, **$p < .01$, ***$p < .001$).}
\label{tab:rep1_rq0_dataset}
\begin{tabular}{lll}
\hline
\textbf{Variables}             & \textbf{Original study} & \textbf{Replication} \\ \hline
(Intercept)                    & 2.81           & 2.47 ***                               \\
test\_file                     & \textbf{1.08 ***}  & \textbf{1.07 ***}                \\
total\_churn                   & \textbf{0.90 ***}   & \textbf{0.90 ***}                 \\
social\_distance               & \textbf{2.35 ***}  & \textbf{2.37 ***}                \\
num\_comments                  & \textbf{0.68 ***}  & \textbf{0.68 ***}                \\
prior\_interaction             & \textbf{1.53 ***}  & \textbf{1.45 ***}                \\
followers\_current             & 1.00                 & 0.98                               \\
main\_team\_member             & \textbf{1.16 ***}  & \textbf{1.19 ***}                \\
age\_current                   & 0.91                 & \textbf{1.01 ***}                \\
team\_size                     & 0.99                 & 0.93                               \\
stars\_current                 & \textbf{0.53 ***}  & \textbf{0.53 ***}                \\
test\_file x num\_comments       & \textbf{1.12 ***}  & \textbf{1.13 ***}                \\
total\_churn x num\_comments     & \textbf{1.06 ***}  & \textbf{1.07 ***}                \\
social\_distance x num\_comments & \textbf{0.92 ***}  & \textbf{0.93 ***}                \\
num\_comments x prior\_inter     & \textbf{1.05 ***}  & \textbf{1.05 ***}                \\ \hline
$R^2m$                         & 0.11                 & 0.11                               \\
$R^2c$                         & 0.58                 & 0.59                               \\
AIC                            & 394,718              & 340,207                            \\ \hline
\end{tabular}
\end{table}

\textit{rq1---Does the personality of a requester affect the likelihood of the pull request being accepted?}

The author of a pull request (a \textit{requester} hereinafter) can  be either a member of a projects' core development team or an outside contributor.
\citeauthor{iyer2019github} examined the personality of requesters to understand whether specific traits  lead to a higher likelihood of pull request acceptance.
As in the original study, we replicate \textit{rq1} using a mixed-effects logistic regression to model the  personality traits of the requester from the new dataset, along with the features already used in \textit{rq0}.

Table~\ref{tab:rep1_rq1_requester} shows the comparison of the odd ratios from our replication against those from the original study.
Unlike \citeauthor{iyer2019github}'s results, in our replication \textit{agreeableness} and  \textit{neuroticism} have a significant effect, respectively positive (1.39) and negative (0.81).
Regarding \textit{openness}, we find that the trait has a positive and significant influence in both studies, albeit the effect is smaller in~\citeauthor{iyer2019github} (1.34 vs. 1.07).
Instead, we find a significant yet opposite effect for both \textit{conscientiousness} (0.77 vs. 1.05) and \textit{extraversion} (1.31 vs. 0.94).

\begin{table}[t]
\caption{Odds ratio of the mixed-effect model with requester's personalities from the original study by \fullcite{iyer2019github} and the replication. Significant results are shown in \textbf{bold} (sig: *$p < .05$, **$p < .01$, ***$p < .001$).}
\label{tab:rep1_rq1_requester}
\begin{tabular}{llc|lc}
\hline
\multirow{3}{*}{\textbf{Variables}} & \multicolumn{2}{c|}{\textbf{Original study}}                                             & \multicolumn{2}{c}{\textbf{Replication}}                               \\
                                    & \multicolumn{1}{c}{\textbf{Single run}} & \multicolumn{1}{c|}{\textbf{Bootstrap.}} & \multicolumn{1}{c}{\textbf{Single run}} & \multicolumn{1}{c}{\textbf{Bootstrap.}} \\ \cline{2-5} 
                                    & \multicolumn{1}{c}{\textbf{Odds Ratio}} & \multicolumn{1}{c|}{\textbf{95\% CI}}       & \multicolumn{1}{c}{\textbf{Odds Ratio}} & \multicolumn{1}{c}{\textbf{95\% CI}}       \\ \hline
(Intercept)                         & 2.87 ***                              & -                                       & 2.72 ***                              & -                                      \\
test\_file                          & \textbf{1.08 ***}                     & [1.05, 1.11]                            & \textbf{1.06 ***}                     & [1.02, 1.10]                           \\
total\_churn                        & \textbf{0.90 ***}                     & [0.89, 0.90]                            & \textbf{0.90 ***}                     & [0.88, 0.91]                           \\
social\_distance                    & \textbf{2.35 ***}                     & [2.59, 2.80]                            & \textbf{2.38 ***}                     & [2.63, 2.99]                           \\
num\_comments                       & \textbf{0.68 ***}                     & [0.65, 0.69]                            & \textbf{0.69 ***}                     & [0.67, 0.69]                           \\
prior\_interaction                  & \textbf{1.52 ***}                     & [1.51, 1.57]                            & \textbf{1.45 ***}                     & [1.43, 1.50]                           \\
followers\_current                  & 0.99                                  & [0.95, 0.98]                            & 0.98                                  & [0.94, 0.99]                           \\
main\_team\_member                  & \textbf{1.15 ***}                     & [1.12, 1.20]                            & \textbf{1.19 ***}                     & [1.15, 1.22]                           \\
age\_current                        & 0.91                                  & [0.83, 0.93]                            & 1.01                                  & [0.98, 1.05]                           \\
team\_size                          & 0.99                                  & [0.85, 0.98]                            & 0.92                                  & [0.84, 0.97]                           \\
stars\_current                      & \textbf{0.54 ***}                     & [0.45, 0.49]                            & \textbf{0.54 ***}                     & [0.46, 0.51]                           \\
openness                            & \textbf{1.07 ***}                     & [1.05, 1.08]                            & \textbf{1.34 ***}                     & [1.31, 1.57]                           \\
conscientiousness                   & \textbf{1.05 ***}                     & [1.03, 1.07]                            & \textbf{0.77 ***}                     & [0.68, 0.79]                            \\
extraversion                        & \textbf{0.94 ***}                     & [0.93, 0.95]                            & \textbf{1.31 ***}                     & [1.29, 1.46]                            \\
agreeableness                       & 1.01                                  & [1.00, 1.02]                            & \textbf{1.39 ***}                     & [1.45, 1.60]                           \\
neuroticism                         & 0.97                                  & [0.94, 0.98]                            & \textbf{0.81 ***}                     & [0.71, 0.84]                           \\
test\_file x num\_comments          & \textbf{1.13 ***}                     & [1.11, 1.15]                            & \textbf{1.12 ***}                     & [1.10, 1.17]                           \\
total\_churn x num\_comments        & \textbf{1.06 ***}                     & [1.06, 1.08]                            & \textbf{1.07 ***}                     & [1.06, 1.09]                           \\
social\_connection x num\_comments  & \textbf{0.92 ***}                     & [0.90, 0.96]                            & \textbf{0.93 ***}                     & [0.90, 0.97]                           \\
num\_comments x prior\_interaction  & \textbf{1.05 ***}                     & [1.06, 1.08]                            & \textbf{1.05 ***}                     & [1.05, 1.08]                           \\ \hline
$R^2m$                              & \multicolumn{2}{c|}{0.11}                                                       & \multicolumn{2}{c}{0.13}                                                        \\
$R^2c$                              & \multicolumn{2}{c|}{0.55}                                                       & \multicolumn{2}{c}{0.59}                                                        \\
AIC                                 & \multicolumn{2}{c|}{394,635}                                                    & \multicolumn{2}{c}{333,552}                                                     \\ \hline
\end{tabular}
\end{table}

\textit{rq2---Does the personality of a closer affect the likelihood of the pull request being accepted?}

Pull requests that are closed by developers (\textit{closers}) who are always part of the core team. 
By analyzing the closers' personalities, \citeauthor{iyer2019github} aimed to understand whether specific traits affect the likelihood of the pull request getting accepted.
As in the original study, we replicate \textit{rq2} by using the requesters' personality traits instead of closers' and modeled them along with the factors used in \textit{rq0}.

The results are reported in Table~\ref{tab:rep1_rq2_closer}.
\textit{Openness} has a significant and positive effect on the pull request acceptance in both the replication (1.18) and the original study (1.05).
We also observe a significant result for \textit{conscientiousness} in both studies, albeit with an opposite direction (0.92 vs. 1.12).
Instead, we find completely contrasting findings regarding \textit{extraversion}, \textit{agreeableness}, and \textit{neuroticism}.

\begin{table}[t]
\caption{Odds ratio of the mixed-effect models with closer's personalities from the original study by \fullcite{iyer2019github} and the replication. Significant results are shown in \textbf{bold} (sig: *$p < .05$, **$p < .01$, ***$p < .001$).}
\label{tab:rep1_rq2_closer}
\begin{tabular}{llc|lc}
\hline
\multirow{3}{*}{\textbf{Variables}} & \multicolumn{2}{c|}{\textbf{Original study}}                        & \multicolumn{2}{c}{\textbf{Replication}}           \\
                                    & \multicolumn{1}{c}{\textbf{Single run}} & \textbf{Bootstrap.} & \multicolumn{1}{c}{\textbf{Single run}} & \textbf{Bootstrap.} \\ \cline{2-5} 
                                    & \multicolumn{1}{c}{\textbf{Odds ratio}} & \textbf{95\% CI}       & \multicolumn{1}{c}{\textbf{Odds ratio}} & \textbf{95\% CI}       \\ \hline
(Intercept)                         & 2.87 ***                              & -                      & 2.45 ***                              & -                      \\
test\_file                          & \textbf{1.08 ***}                     & [1.05, 1.11]           & \textbf{1.06 ***}                     & [1.03, 1.10]            \\
total\_churn                        & \textbf{0.90 ***}                     & [0.89, 0.91]           & \textbf{0.90 ***}                     & [0.88, 0.90]            \\
social\_distance                    & \textbf{2.35 ***}                     & [2.35, 2.83]           & \textbf{2.41 ***}                     & [2.66, 3.03]            \\
num\_comments                       & \textbf{0.68 ***}                     & [0.65, 0.68]           & \textbf{0.69 ***}                     & [0.67, 0.69]            \\
prior\_interaction                  & \textbf{1.49 ***}                     & [1.52, 1.56]           & \textbf{1.45 ***}                     & [1.43, 1.50]            \\
followers\_current                  & 0.98                                  & [0.94, 0.99]           & 0.98                                  & [0.94, 0.99]            \\
main\_team\_member                  & \textbf{1.16 ***}                     & [1.12, 1.21]           & \textbf{1.19 ***}                     & [1.14, 1.22]            \\
age\_current                        & \textbf{0.92 ***}                     & [0.86, 0.95]           & 1.01                                  & [0.98, 1.05]            \\
team\_size                          & 0.97                                  & [0.88, 1.00]           & 0.97                                  & [0.88, 1.00]            \\
stars\_current                      & \textbf{0.54 ***}                     & [0.44, 0.51]           & \textbf{0.54 ***}                     & [0.46, 0.50]            \\
openness                            & \textbf{1.05 *}                       & [1.02, 1.10]           & \textbf{1.18 ***}                     & [1.15, 1.27]            \\
conscientiousness                   & \textbf{1.12 ***}                     & [1.11, 1.18]           & \textbf{0.92 ***}                     & [0.86, 0.95]            \\
extraversion                        & \textbf{1.06 ***}                     & [1.06, 1.13]           & 1.03                                  & [0.99, 1.05]            \\
agreeableness                       & 1.01                                  & [0.99, 1.04]           & \textbf{1.13 ***}                     & [1.13, 1.21]            \\
neuroticism                         & \textbf{1.08 ***}                     & [1.06, 1.14]           & 1.00                                  & [0.94, 1.06]            \\
test\_file x num\_comments          & \textbf{1.12 ***}                     & [1.08, 1.16]           & \textbf{1.13 ***}                     & [1.10, 1.17]            \\
total\_churn x num\_comments        & \textbf{1.06 ***}                     & [1.05, 1.08]           & \textbf{1.07 ***}                     & [1.06, 1.09]            \\
social\_connection x num\_comments  & \textbf{0.92 ***}                     & [0.88, 0.95]           & \textbf{0.93 ***}                     & [0.90, 0.97]            \\
num\_comments x prior\_interaction  & \textbf{1.05 ***}                     & [1.05, 1.08]           & \textbf{1.05 ***}                     & [1.05, 1.07]            \\ \hline
$R^2m$                              & \multicolumn{2}{c|}{0.12}                                      & \multicolumn{2}{c}{0.11}                                         \\
$R^2c$                              & \multicolumn{2}{c|}{0.57}                                      & \multicolumn{2}{c}{0.59}                                         \\
AIC                                 & \multicolumn{2}{c|}{394,548}                                   & \multicolumn{2}{c}{333,965}                                      \\ \hline
\end{tabular}
\end{table}

\textit{rq3---Does the difference in personality between the requester and the closer affect the likelihood of the pull request being accepted?}

Finally, \citeauthor{iyer2019github} analyzed the differences in the personality traits between the requester and the closer to understand whether they  hinder or facilitate pull request acceptance.
As in the original study, we replicate \textit{rq3} by considering the effects of personality differences in the model by adding the absolute differences between the personality traits of requesters and closers along with the other socio-technical features used in \textit{rq0}.

The results are reported in Table~\ref{tab:rep1_rq3_diff}.
Regarding \textit{openness}, the difference  between the requester and the closer is positive and significant (1.07) only in the replication.
We observe consistent results regarding the difference in the levels of \textit{conscientiousness} and \textit{extraversion}, which have a positive effect in both studies.
Instead, we observe contrasting results for the difference in the levels of \textit{agreeableness} and \textit{neuroticism}---negative in the replication (respectively, 0.94 and 0.93) and positive in the original study (1.02 and 1.22).

\begin{table}[t]
\caption{Odds ratio of the mixed-effect model with personality differences between the requester and closer from the original study by \fullcite{iyer2019github} and the replication. Significant results are shown in \textbf{bold} (sig: *$p < .05$, **$p < .01$, ***$p < .001$).}
\label{tab:rep1_rq3_diff}
\begin{tabular}{llc|lc}
\hline
\multirow{3}{*}{\textbf{Variables}} & \multicolumn{2}{c|}{\textbf{Original study}} & \multicolumn{2}{c}{\textbf{Replication}}        \\
                                    & \textbf{Single run} & \textbf{Bootstrap.} & \multicolumn{1}{c}{\textbf{Single run}} & \textbf{Bootstrap.} \\ \cline{2-5} 
                                    & \textbf{Odds ratio} & \textbf{95\% CI}    & \multicolumn{1}{c}{\textbf{Odds ratio}} & \textbf{95\% CI}    \\ \hline
(Intercept)                         & 3.34 ***            & -                   & 2.68 ***            & -                   \\
test\_file                          & \textbf{1.09 ***}   & [1.05, 1.12]        & \textbf{1.09 ***}   & [1.05, 1.12]        \\
total\_churn                        & \textbf{0.92 ***}   & [0.90, 0.93]        & \textbf{0.90 ***}   & [0.89, 0.91]        \\
social\_distance                    & \textbf{1.81 ***}   & [1.86, 2.03]        & \textbf{2.34 ***}   & [2.55, 2.91]        \\
num\_comments                       & \textbf{0.66 ***}   & [0.65, 0.67]        & \textbf{0.68 ***}   & [0.67, 0.68]        \\
prior\_interaction                  & \textbf{1.66 ***}   & [1.63, 1.69]        & \textbf{1.48 ***}   & [1.46, 1.53]        \\
followers\_current                  & \textbf{1.07 ***}   & [1.06, 1.11]        & \textbf{1.07 ***}   & [1.06, 1.11]        \\
main\_team\_member                  & \textbf{1.27 ***}   & [1.23, 1.31]        & \textbf{1.22 ***}   & [1.17, 1.25]        \\
age\_current                        & \textbf{0.92 ***}   & [0.87, 0.93]        & 1.01                & [0.95, 1.05]        \\
team\_size                          & 0.96                & [0.89, 1.00]        & 0.94                & [0.87, 0.99]        \\
stars\_current                      & \textbf{0.55 ***}   & [0.44, 0.50]        & \textbf{0.53 ***}   & [0.46, 0.49]        \\
diff\_openness\_abs                 & 1.01                & [1.01, 1.04]        & \textbf{1.07 ***}   & [1.06, 1.14]        \\
diff\_conscientiousness\_abs        & \textbf{1.29 ***}   & [1.29, 1.35]        & \textbf{1.25 ***}   & [1.28, 1.34]        \\
diff\_extraversion\_abs             & \textbf{1.12 ***}   & [1.11, 1.16]        & \textbf{1.31 ***}   & [1.32, 1.44]        \\
diff\_agreeableness\_abs            & \textbf{1.02 **}    & [1.00, 1.04]        & \textbf{0.94 ***}   & [0.89, 0.95]        \\
diff\_neuroticism\_abs              & \textbf{1.22 ***}   & [1.21, 1.27]        & \textbf{0.93 ***}   & [0.88, 0.94]        \\
test\_file x num\_comments          & \textbf{1.11 ***}   & [1.09, 1.15]        & \textbf{1.13 ***}   & [1.10, 1.17]        \\
total\_churn x num\_comments        & \textbf{1.06 ***}   & [1.05, 1.07]        & \textbf{1.07 ***}   & [1.06, 1.09]        \\
social\_connection x num\_comments  & \textbf{0.93 ***}   & [0.89, 0.97]        & \textbf{0.93 ***}   & [0.90, 0.98]        \\
num\_comments x prior\_interaction  & \textbf{1.05 ***}   & [1.05, 1.08]        & \textbf{1.05 ***}   & [1.05, 1.07]        \\ \hline
$R^2m$                              & \multicolumn{2}{c|}{0.13}                 & \multicolumn{2}{c}{0.12}                                      \\
$R^2c$                              & \multicolumn{2}{c|}{0.56}                 & \multicolumn{2}{c}{0.58}                                      \\
AIC                                 & \multicolumn{2}{c|}{390,495}              & \multicolumn{2}{c}{333,114}                                   \\ \hline
\end{tabular}
\end{table}

\subsubsection{Replication of \citeauthor{calefato2019apache}~(\citeyear{calefato2019apache})} 

In this section, we answer the \revone{six} research questions \textit{rq1-\revone{6}} presented in the original study by \fullcite{calefato2019apache}. In the replication, we recreate  the related figures and tables for comparison.

\citeauthor{calefato2019apache} conducted a preliminary analysis to rule out changes in personality over time. 
For each of the $N=211$ developers in the dataset, they computed monthly-based personality scores, then split the set by date into two subsets of approximately the same size. 
For each trait, they averaged the scores in each subset, thus obtaining two observations for each developer (i.e., early vs. later). 
Finally, for each trait, they performed a Wilcoxon Signed-Rank test to verify the null hypothesis that the median difference between pairs of observations (i.e., for each developer) was not significantly different from zero. Table~\ref{tab:rep2_rq0_time_changes} reports the results from the five paired tests in both the original study and the replication.
We replicate the same tests after replacing the original dataset, containing personality scores obtained from \pin, with the new dataset, containing the scores obtained using \liwc.
The results show no significant differences between the distributions (all adjusted p-values > 0.05 after Bonferroni correction for multiple tests), thus confirming the stability of personality traits over time with both personality tools.

\begin{table}[t]
\caption{Results of the Wilcoxon Signed-Rank tests for assessing changes in mean personality traits over time in the original study by \fullcite{calefato2019apache} and the replication (all p-values > 0.05 after Bonferroni correction)}
\label{tab:rep2_rq0_time_changes}
\begin{tabular}{lccc|l|cccl}
\hline
\multicolumn{1}{c}{\multirow{2}{*}{\textbf{Trait}}} &
  \multicolumn{4}{c|}{\textbf{Original study}} &
  \multicolumn{4}{c}{\textbf{Replication}} \\ \cline{2-9} 
\multicolumn{1}{c}{} &
  \textbf{V} &
  \textbf{p-value} &
  \multicolumn{2}{c|}{\textbf{95\% CI}} &
  \textbf{V} &
  \textbf{p-value} &
  \multicolumn{2}{c}{\textbf{95\% CI}} \\ \hline
Openness          & 6,109 & 0.589 & \multicolumn{2}{c|}{[-0.002, -0.003]} & 9,330 & 1,000 & \multicolumn{2}{c}{[-0.006, 0.029]} \\
Conscientiousness & 5,575 & 0.661 & \multicolumn{2}{c|}{[-0.004, -0.003]} & 8,320 & 1,000 & \multicolumn{2}{c}{[-0.014, 0.014]} \\
Extraversion       & 5,839 & 0.964 & \multicolumn{2}{c|}{[-0.003, -0.003]} & 7,751 & 1,000 & \multicolumn{2}{c}{[-0.022, 0.008]} \\
Agreeableness      & 5,871 & 0.917 & \multicolumn{2}{c|}{[-0.003, -0.003]} & 7,199 & 0.448 & \multicolumn{2}{c}{[-0.028, 0.002]} \\
Neuroticism        & 5,915 & 0.853 & \multicolumn{2}{c|}{[-0.003, -0.004]} & 9,075 & 1,000 & \multicolumn{2}{c}{[-0.008, 0.023]} \\ \hline
\end{tabular}
\end{table}

\textit{rq1---Are there groupings of similar developers according to their personality profile?}

To answer the first research question, \citeauthor{calefato2019apache} applied several techniques  to reveal the presence of natural groupings of personalities within the dataset of $N=211$ developers.
We replicate the same analyses presented in \fullcite{calefato2019apache} on the new dataset.

First, to ensure that original data was suitable for structure detection, \citeauthor{calefato2019apache} computed the Kaiser-Meyer-Olkin measure (0.5, the minimum value recommended in literature~\cite{field2012kmo}) and Barlett's test of sphericity ($\chi^2$ = 4088.32, p < 0.001). 
We obtain similar results with the new dataset (KMO = 0.5; $\chi^2$ = 900, p < 0.001).
Accordingly, we proceed with the analyses to uncover latent factors.  

The first analysis performed was the Principal Component Analysis (PCA), a statistical procedure that converts a set of observations of possibly correlated variables into a set of values of linearly uncorrelated variables, i.e., the principal components.

\begin{figure}[tb]
\centering
    \subfloat[Original study\label{fig:rep2_rq1_pca_screeplot_orig}]{%
       \includegraphics[width=0.4\textwidth]{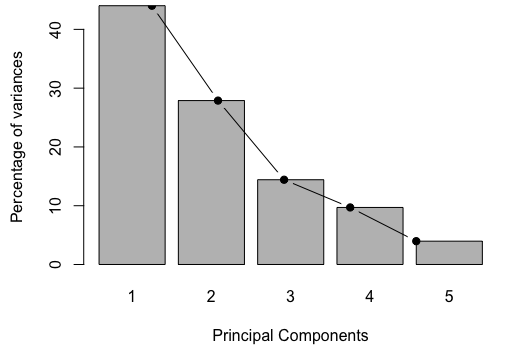}
     }
    \subfloat[Replication\label{fig:rep2_rq1_pca_screeplot_repl}]{%
       \includegraphics[width=0.5\textwidth]{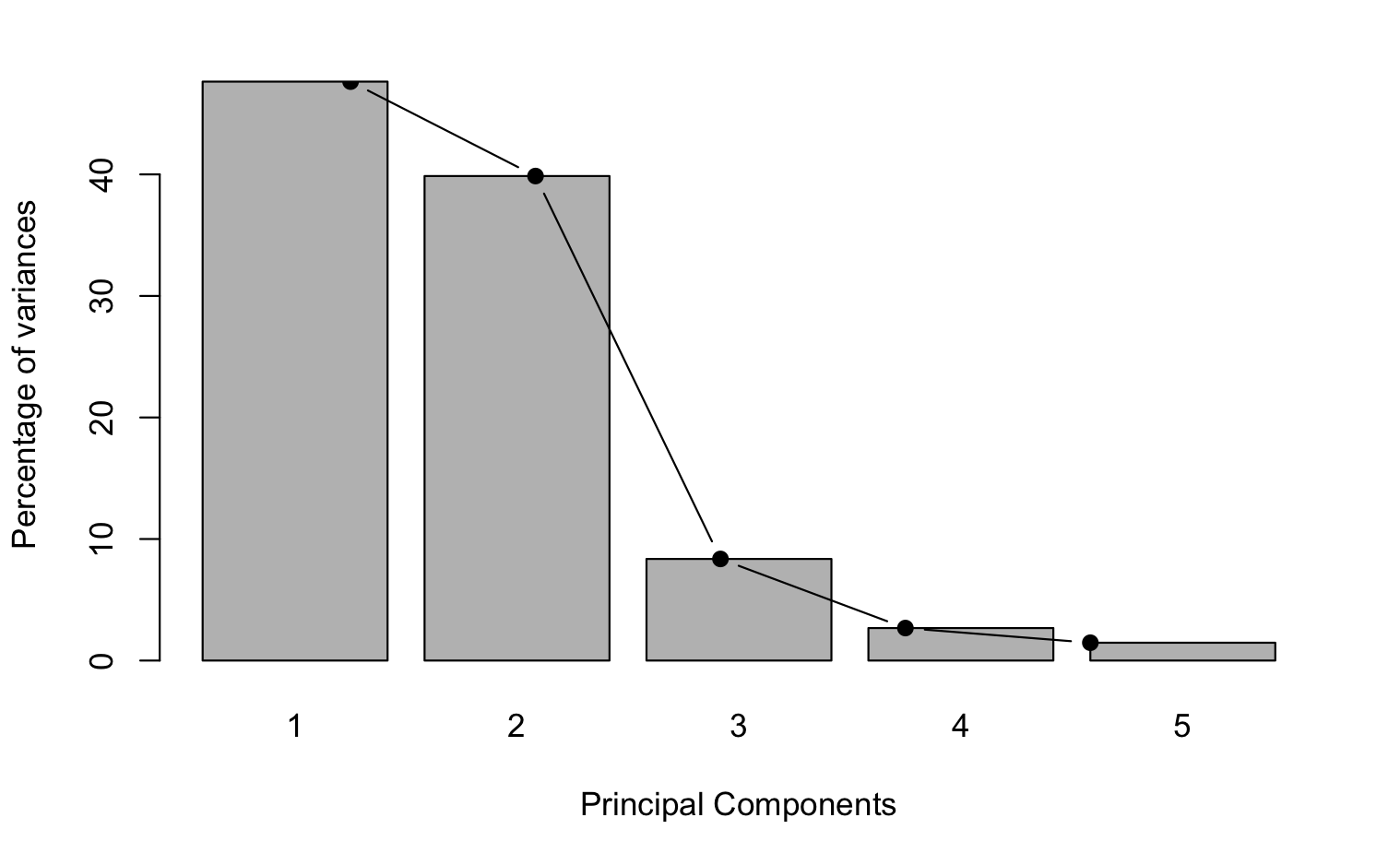}
     }
\caption{Percent of variance explained by the principal components in the original study by \fullcite{calefato2019apache} (a) and the replication (b).}
\label{fig:rep2_rq1_pca_screeplots}
\end{figure}

The scree plots in Fig.~\ref{fig:rep2_rq1_pca_screeplots} show the percentage of variance in the data for each of the five components extracted from the data. 
In the original study, the first three components accounted for most of the variance in the data (86\%, see Fig.~\ref{fig:rep2_rq1_pca_screeplot_orig}), whereas in the replication the first two account for nearly 88\% (Fig.~\ref{fig:rep2_rq1_pca_screeplot_repl}).
The analysis of the eigenvalues in Table~\ref{tab:rep2_rq1_pca_components} shows that only the first two components in each replication have a value over Kaiser's criterion of 1, the cut-off point typically used to retain principal components.
Eigenvalues correspond to the amount of the variation explained by each principal component; 
therefore, a latent component has an eigenvalue > 1 when it accounts for more variance than its accounted for by the original variables in a dataset.
Next, we check how the traits load on the two extracted components.
The loadings from the original study and the replication are reported in Table~\ref{tab:rep2_rq1_pca_loadings}, from which we observe inconsistent results.
In the original study, \textit{openness} and \textit{neuroticism} load on the first component, whereas \textit{conscientiousness}, \textit{extraversion}, and \textit{agreeableness} on the second. 
Conversely, in the replication \textit{openness}, \textit{extraversion}, and \textit{agreeableness} load on the first component, whereas  \textit{conscientiousness} and \textit{neuroticism} load on the second;
we also observe that \textit{openness} and \textit{conscientiousness} load negatively on their respective component.

\begin{table}[tb]
\caption{Eigenvalues returned by PCA in the original study and the replication. Only the components in \textbf{bold} with eigenvalue \textgreater 1 are retained.}
\label{tab:rep2_rq1_pca_components}
\begin{tabular}{lcc|cc}
\hline
 &
  \multicolumn{2}{c|}{\textbf{Original study}} &
  \multicolumn{2}{c}{\textbf{Replication}} \\ \cline{2-5} 
 &
  \multicolumn{1}{c}{\textbf{Eigenvalue}} &
  \multicolumn{1}{c|}{\textbf{\% of variance}} &
  \multicolumn{1}{c}{\textbf{Eigenvalue}} &
  \multicolumn{1}{c}{\textbf{\% of variance}} \\ \hline
\textbf{Component 1} & \textbf{2.201} &  44.023  &  \textbf{2.382} &  47.63  \\
\textbf{Component 2} & \textbf{1.394} &  27.885  &  \textbf{1.993} &  39.86  \\ \hline
Component 3          &  0.721         &  14.419  &  0.418          &   8.36  \\
Component 4          &  0.485         &   9.705  &  0.134          &   2.67  \\
Component 5          &  0.198         &   3.967  &  0.073          &   1.47  \\ \hline
\end{tabular}
\end{table}

\begin{table}[tb]
\caption{Standardized loadings for the extracted principal components in the original study by \fullcite{calefato2019apache} and the replication.}
\label{tab:rep2_rq1_pca_loadings}
\begin{tabular}{lcc|cc}
\hline
\multicolumn{1}{c}{\multirow{2}{*}{\textbf{Trait}}} & \multicolumn{2}{c|}{\textbf{Original study}} & \multicolumn{2}{c}{\textbf{Replication}}    \\ \cline{2-5} 
\multicolumn{1}{c}{}                                & \textbf{Component 1}         & \textbf{Component 2}         & \textbf{Component 1} & \textbf{Component 2} \\ \hline
Openness          &  \textbf{0.79} &  0.03           &  \textbf{-0.861} &  0.414  \\
Conscientiousness &  \textbf{0.69} &  0.44           &  0.118           &  \textbf{0.942}  \\
Extraversion      &  0.27          &  \textbf{0.74}  &  \textbf{0.849}  &  0.132  \\
Agreeableness     &  -0.15         &  \textbf{0.92}  &  \textbf{0.867 } &  0.357  \\
Neuroticism       &  \textbf{0.89} &  0.04           &  0.001           &  \textbf{-0.971}  \\ \hline
\end{tabular}
\end{table}

After applying PCA, \citeauthor{calefato2019apache} applied the \textit{k}-means clustering algorithm to extract clusters of developers' personalities.
We replicate the analysis on the new dataset and use the `elbow' method to identify the optimal number of clusters. 
The elbow point corresponds to the smallest \textit{k} value after which is not observed a large decrease in the within-group heterogeneity---measured using the sum of squares---with the increase of the number of clusters.
The scree plots from both studies are reported in Appendix~\ref{appendix:additional-material-ph2}, Fig.~\ref{fig:rep2_rq1_cluster_screeplots}. 
In the original study, \citeauthor{calefato2019apache} selected \textit{k = 3}, whereas we choose \textit{k=2}.
Table~\ref{tab:rep2_rq1_kmeans_clusters} shows the distribution of the developers across the personality clusters extracted in the two studies. 
The developers are fairly evenly distributed across the clusters in the original study. 
In the replication, the first cluster is twice the size of the second, though we observed an even larger imbalance with larger \textit{k} values.
The table also reports the coordinates of the centroids---the average position of the elements assigned to a cluster. 
All the values are z-score standardized, with positive (negative) values above (below) the overall means.
Then, \citeauthor{calefato2019apache} performed five non-parametric Kruskal-Wallis tests (one per trait), to understand whether the trait distributions in the clusters are significantly different, followed by a Tukey and Kramer (Nemenyi) \textit{post hoc} test for multiple pairwise comparisons, to understand which pairs are indeed different.
Table~\ref{tab:rep2_rq1_kw} shows the results of the Kruskal-Wallis tests, after applying Bonferroni correction to the p-values for repeated tests.
In the original study, the Kruskal-Wallis test for each of the five traits was statistically significant (p < 0.001) with a relatively strong ($\epsilon^2 \geq 0.16$) or strong effect size ($\epsilon^2 \geq 0.36$)~\cite{rea2014epsilonsquared}; however the \textit{post hoc} tests showed that only clusters 1 \& 3 and clusters 2 \& 3 were significantly different from each other.
In the replicated study, the Kruskal-Wallis tests are also all significant (p < 0.001), with the effect size ranging between moderate ($\epsilon^2 \geq 0.04$) and strong.
Because there are only two clusters, there is no need to run a \textit{post hoc} test to affirm that there are significant differences among the trait distributions.

\begin{table}[tb]
\caption{{Size and centers of the three clusters extracted with \textit{k}-means in the original study by \fullcite{calefato2019apache} and the replication. The highest $\blacktriangle$ and lowest $\blacktriangledown$ values per trait are shown in \textbf{bold}.} }
\label{tab:rep2_rq1_kmeans_clusters}
\def\arraystretch{1}
\ignorespaces 
\centering 
\begin{tabular}{lccccc}
\multicolumn{6}{c}{\textbf{Original study}}\\
\hline 
\textbf{Cluster (size)}& \textbf{Openness} & \textbf{Conscient.} & \textbf{Extraver.} & \textbf{Agreeabl.} & \textbf{Neurotic.}\\
\hline 
Cluster 1 (76) &
  \textbf{-0.74}$\blacktriangledown$ &
  \textbf{-0.69}$\blacktriangledown$ &
  -0.06 &
  0.37 &
  \textbf{-0.84}$\blacktriangledown$\\
Cluster 2 (55) &
  \textbf{0.90}$\blacktriangle$ &
  \textbf{0.86}$\blacktriangle$ &
  \textbf{0.99}$\blacktriangle$ &
  \textbf{0.45}$\blacktriangle$ &
  \textbf{0.81}$\blacktriangle$\\
Cluster 3 (80) &
  0.08 &
  0.07 &
  \textbf{-0.62}$\blacktriangledown$ &
  \textbf{-0.67$\blacktriangledown$} &
  0.25\\
\hline 
\multicolumn{6}{l}{}\\
\multicolumn{6}{c}{\textbf{Replication}}\\
\hline 
\textbf{Cluster (size)}& \textbf{Openness} & \textbf{Conscient.} & \textbf{Extraver.} & \textbf{Agreeabl.} & \textbf{Neurotic.}\\
\hline 
Cluster 1 (156) &
  \textbf{0.50}$\blacktriangle$ &
  \textbf{0.16}$\blacktriangle$ &
  \textbf{-0.36}$\blacktriangledown$ &
  \textbf{-0.32}$\blacktriangledown$ &
  \textbf{-0.22}$\blacktriangledown$\\
Cluster 2 (76) &
  \textbf{-1.03}$\blacktriangledown$ &
  \textbf{-0.33}$\blacktriangledown$ &
  \textbf{0.77}$\blacktriangle$ &
  \textbf{0.65}$\blacktriangle$ &
  \textbf{0.45}$\blacktriangle$\\
\hline 
\end{tabular}\par 
\end{table}

\begin{table}[tb]
\caption{Results of the Kruskal-Wallis tests for the comparisons of the distributions of each personality trait scores across the clusters in the original study by \fullcite{calefato2019apache} and the replication. All p-values < 0.001 after Bonferroni correction.}
\label{tab:rep2_rq1_kw}
\begin{tabular}{lcccc|cccc}
\hline
\multirow{2}{*}{\textbf{Trait}} & \multicolumn{4}{c|}{\textbf{Original study}}                                         & \multicolumn{4}{c}{\textbf{Replication}}                                                            \\ \cline{2-9} 
                                & \textbf{$\chi^2$} & \textbf{p-value} & \textbf{\ensuremath{\epsilon^2}} & \textbf{95\% CI} & \textbf{$\chi^2$} & \textbf{p-value} & \textbf{\ensuremath{\epsilon^2}} & \textbf{95\% CI} \\ \hline
Openness                        & 87.836               & \textless 0.001  & 0.418                                  & [0.297, 0.532]   & 136                  & \textless 0.001  & 0.590                                  & [0.505, 0.658]   \\
Conscienti.               & 78.777               & \textless 0.001  & 0.375                                  & [0.257, 0.495]   & 17                   & \textless 0.001  & 0.073                                  & [0.021, 0.149]   \\
Extraversion                    & 94.554               & \textless 0.001  & 0.450                                  & [0.354, 0.547]   & 84                   & \textless 0.001  & 0.362                                  & [0.264, 0.463]   \\
Agreeablen.                   & 61.248               & \textless 0.001  & 0.292                                  & [0.197, 0.401]   & 62                   & \textless 0.001  & 0.270                                  & [0.170, 0.377]   \\
Neuroticism                     & 107.560              & \textless 0.001  & 0.512                                  & [0.410, 0.613]   & 24                   & \textless 0.001  & 0.104                                  & [0.041, 0.185]   \\ \hline
\end{tabular}
\end{table}

Finally, \citeauthor{calefato2019apache} applied Archetypal Analysis.
In Appendix~\ref{appendix:additional-material-ph2}, Fig.~\ref{fig:rep2_rq1_archetypes_screeplots} we report the scree plots used to identify the optimal number of archetypes to extract with the elbow criterion. 
The two plots show the fraction of total variance in the data explained by the number of extracted archetypes.
In the original study, \citeauthor{calefato2019apache} extracted three archetypes; in the replication, the function also plateaus after three archetypes.
Therefore, also for the ease of comparison with the original study, we opt for extracting three archetypes.
Table~\ref{tab:rep2_rq1_archetypes} shows the trait coordinates, standardized for ease of comparison, for the archetypes extracted in the two studies. 
Looking at the trait coordinates, the archetypal analyses in the two studies do not extract similar phenotypes of developers' personalities.

\begin{table}[tb]
\caption{{The archetypes extracted in the original study by \fullcite{calefato2019apache} and the replication. The highest $\blacktriangle$ and lowest $\blacktriangledown$ standardized values per trait are shown in \textbf{bold}.} }
\label{tab:rep2_rq1_archetypes}
\def\arraystretch{1}
\ignorespaces 
\centering 
\begin{tabular}{cccccc}
\multicolumn{6}{c}{\textbf{Original study}}\\
\hline 
\textbf{Archetype} & \textbf{Openness} & \textbf{Conscient.} & \textbf{Extraver.} & \textbf{Agreeabl.} & \textbf{Neuroticism}\\
\hline 
Archetype 1 &
  0.51 &
  -0.13 &
  \textbf{-0.81}$\blacktriangledown$ &
  \textbf{-1.09}$\blacktriangledown$ &
  \textbf{0.61}$\blacktriangle$\\
Archetype 2 &
  \textbf{0.64}$\blacktriangle$ &
  \textbf{1.06}$\blacktriangle$ &
  \textbf{1.12}$\blacktriangle$ &
  \textbf{0.87}$\blacktriangle$ &
  0.54\\
Archetype 3 &
  \textbf{-1.15}$\blacktriangledown$ &
  \textbf{-0.93 $\blacktriangledown$} &
  -0.31 &
  0.23 &
  \textbf{-1.15}$\blacktriangledown$\\
\hline 
\multicolumn{6}{l}{}\\
\multicolumn{6}{c}{\textbf{Replication}}\\
\hline 
\textbf{Archetype} & \textbf{Openness} & \textbf{Conscient.} & \textbf{Extraver.} & \textbf{Agreeabl.} & \textbf{Neuroticism}\\
\hline 
Archetype 1 &
  \textbf{0.32}$\blacktriangledown$ &
  \textbf{1.30}$\blacktriangle$ &
  \textbf{-0.68}$\blacktriangle$ &
  \textbf{0.35}$\blacktriangle$ &
  \textbf{-1.28}$\blacktriangledown$\\
Archetype 2 &
  \textbf{1.10}$\blacktriangle$ &
  \textbf{0.72}$\blacktriangledown$ &
  \textbf{-1.36}$\blacktriangledown$ &
  \textbf{-0.62}$\blacktriangledown$ &
  \textbf{ 0.17}$\blacktriangle$\\
Archetype 3 &
  1.08 &
  1.03 &
 -1.01 &
 -0.31 &
 -0.80\\
\hline 
\end{tabular}\par 
\end{table}

\textit{rq2---Do developers' personality traits vary with the type of contributors (i.e., core vs. peripheral?)}

In the second research question, \citeauthor{calefato2019apache} investigated whether the members of projects' core development teams exhibit different personality traits.
Accordingly, they first filtered the personality scores retaining only the $N=118$ commit authors and then split this set into two subgroups, namely peripheral developers (i.e., external contributors without commit access to the repositories, $N=62$) and core developers (i.e., project members with write access to the source code repository, $N=56$). 
To replicate the analysis, we perform for each trait a Wilcoxon Rank Sum test for unpaired group comparison on the dataset with the new personality scores.

The results from both the original study and the replication are reported in Table~\ref{tab:rep2_rq2_wilcoxon_core_vs_per}.
Consistently, in both studies, we observe no significant differences (i.e., all adjusted p-values $> 0.05$, after Bonferroni correction) across all five traits.
As such, the two studies consistently find that, on average, the personalities of core developers are not significantly different from those of peripheral developers.

\begin{table}[tb]
\caption{Results of the Wilcoxon Rank Sum tests for the unpaired comparison of median personality trait scores between core and peripheral developers in the original study by \fullcite{calefato2019apache} and the replication.  All p-values {\textgreater} 0.05 after Bonferroni correction.}
\label{tab:rep2_rq2_wilcoxon_core_vs_per}
\begin{tabular}{lccc|ccc}
\hline
\multicolumn{1}{c}{\multirow{2}{*}{\textbf{Trait}}} & \multicolumn{3}{c|}{\textbf{Original study}} & \multicolumn{3}{c}{\textbf{Replication}}         \\ \cline{2-7} 
\multicolumn{1}{c}{}                                & \textbf{W}     & \textbf{p-value}     & \textbf{95\% CI}    & \textbf{W} & \textbf{p-value} & \textbf{95\% CI} \\ \hline
Openness                                            &  1,583         &  1.000               &  [-0.009, 0.008]    & 2,233      & 0.500            & [-0.009, 0.087]  \\
Conscientiousness                                   &  1,625         &  1.000               &  [-0.010, 0.011]    & 1,989      & 1.000            & [-0.022, 0.034]  \\
Extraversion                                        &  1,575         &  1.000               &  [-0.010, 0.008]    & 1,902      & 1.000            & [-0.041, 0.041]  \\
Agreeableness                                       &  1,273         &  0.271               &  [-0.017, 0.000]    & 1,685      &  1.000           & [-0.064, 0.018]  \\
Neuroticism                                         &  2,051         &  0.063               &  [0.004, 0.027]     & 1,751      &  1.000           & [-0.049, 0.021]  \\ \hline
\end{tabular}
\end{table}

\textit{rq3---Do developers' personality traits change after becoming a core member of a project's development team?}

In the third research question, \citeauthor{calefato2019apache} investigated whether developers exhibit different personality traits after becoming members of a project's core development team.
Accordingly, for each of the $N=56$ core developers with write access to source code repositories, they first retrieved the date of the first commit accepted and integrated by them, as an approximation of the moment when they have become a member of a project's core development team. 
Then, for any of the projects they gained membership for, they used that date to split the personality trait scores of the developers into two paired groups, i.e., before vs. after becoming a project’s core team member. 
We replicate the same analyses on the new dataset.

In Appendix~\ref{appendix:additional-material-ph2}, Fig.~\ref{fig:rep2_rq3_boxplot_membership} we report the boxplots of the five personality scores across the two groups in both the original study and the replication.
Also, Table~\ref{tab:rep2_rq3_wsr-membership} reports the results of the five Wilcoxon Signed- Rank tests executed (one per trait). 
No significant differences are returned by the tests in both studies (all adjusted p-values $> 0.05$ after Bonferroni correction).
As such, the two studies consistently find that developers' personality does not change significantly after becoming a core developer.

\begin{table}[tb]
\caption{Results of the Wilcoxon Signed-Rank tests for the paired comparison of mean personality trait scores of developers before vs. after becoming members of a project's core team in the original study by \fullcite{calefato2019apache} and the replication.  All p-values $> 0.05$ after Bonferroni correction.}
\label{tab:rep2_rq3_wsr-membership}
\begin{tabular}{lccc|ccc}
\hline
\multicolumn{1}{c}{\multirow{2}{*}{\textbf{Trait}}} & \multicolumn{3}{c|}{\textbf{Original study}} & \multicolumn{3}{c}{\textbf{Replication}}         \\ \cline{2-7} 
\multicolumn{1}{c}{}                                & \textbf{V}     & \textbf{p-value}     & \textbf{95\% CI}    & \textbf{V} & \textbf{p-value} & \textbf{95\% CI} \\ \hline
Openness                                            &  39            &  1.000               &  [-0.011, 0.034]    &  62 &  1.000 &  [-0.074, 0.141]                  \\
Conscientiousness                                   &  40            &  1.000               &  [-0.008, 0.031]    &  76 &  0.765 &  [-0.014, 0.112]                  \\
Extraversion                                        &  17            &  1.000               &  [-0.019, 0.019]    &  46 &  1.000 &  [-0.126, 0.071]                   \\
Agreeableness                                       &  15            &  1.000               &  [-0.038, 0.011]    &  49  &  1.000 &  [-0.121, 0.081]                  \\
Neuroticism                                         &  43            &  0.654               &  [-0.005, 0.048]    & 35  & 1.000  & [-0.118, 0.018]                  \\ \hline
\end{tabular}
\end{table}

\textit{rq4---Do developers' personality traits vary with the degree of development activity?}

\citeauthor{calefato2019apache} investigated whether more productive developers are characterized by specific personality trait levels.
Using the same groups of core ($N=56$) and peripheral ($N=62$) developers created earlier for \textit{rq2}, they further split the two sets according to the level of development activity.
Specifically, they found the mean number of commits authored by each developer in the peripheral group and split it into two subsets, i.e., authored-commits-high ($N=17$) and authored-commits-low ($N=45$).
Similarly, they created the integrated-commits-high ($N=44$) and integrated-commits-low ($N=8$) subgroups considering the mean number of commits integrated (i.e., accepted) by the developers in the core group. 
We replicate this research question on the new dataset by applying a series of Wilcoxon Rank Sum tests to make unpaired comparisons of the median personality scores between high- and low-activity developers. 

The results from the original study and the replications are shown in Table~\ref{tab:rep2_rq4_wrs_high_low}. 
The test results reveal no cases of statistically significant differences between the pairs of trait distributions in both studies (i.e., adjusted p-values $> 0.05$ after Bonferroni correction).
As such, the two studies consistently find that developers' personality does not vary significantly with the level of development activity.

\begin{table}[t]
\caption{{Results of the Wilcoxon Rank Sum test for the unpaired comparison of median personality trait scores between developers with high vs. low degree of development activity in the original study by \fullcite{calefato2019apache} and the replication. All p-values $> 0.05$ after Bonferroni correction.} }
\label{tab:rep2_rq4_wrs_high_low}
\def\arraystretch{1}
\ignorespaces 
\centering 
\begin{tabular}{llccc}
\multicolumn{5}{c}{\textbf{Original study}}\\
\hline 
 & \textbf{Trait} & \textbf{W} & \textbf{p-value} & \textbf{95\% CI}\\
\hline 
 &
  Openness &
  476 &
  1.000 &
  [-0.004, 0.021]\\
High vs. low &
  Conscientiousness &
  449 &
  1.000 &
  [-0.008, 0.024]\\
commit authors &
  Extraversion &
  383 &
  1.000 &
  [-0.017, 0.017]\\
(peripheral devs) &
  Agreeableness &
  341 &
  1.000 &
  [-0.018, 0.009]\\
 &
  Neuroticism &
  408 &
  1.000 &
  [-0.013, 0.017]\\
  \hline
 &
  Openness &
  193 &
  1.000 &
  [-0.014, 0.020]\\
High vs. low &
  Conscientiousness &
  163 &
  1.000 &
  [-0.029, 0.019] \\
commit integrators &
  Extraversion &
  129 &
  1.000 &
  [-0.028, 0.006] \\
(core devs) &
  Agreeableness &
  204 &
  1.000 &
  [-0.013, 0.025] \\
 &
  Neuroticism &
  151 &
  1.000 &
  [-0.040, 0.017]\\
  \hline
  
\multicolumn{5}{l}{}\\
\multicolumn{5}{c}{\textbf{Replication}}\\
\hline 
 & \textbf{Trait} & \textbf{W} & \textbf{p-value} & \textbf{95\% CI}\\
\hline 
 &
  Openness &
  520 &
  1.000 &
  [-0.039, 0.122]\\
High vs. low &
  Conscientiousness &
  557 &
  1.000 &
  [-0.021, 0.078]\\
commit authors &
  Extraversion &
  520 &
  1.000 &
  [-0.115, 0.041]\\
(peripheral devs) &
  Agreeableness &
  489 &
  1.000 &
  [-0.078, 0.100]\\
 &
  Neuroticism &
  389 &
  1.000 &
  [-0.084, 0.035]\\
  \hline
 &
  Openness &
  165 &
  1.000 &
  [-0.139, 0.046]\\
High vs. low &
  Conscientiousness &
  253 &
  1.000 &
  [-0.035, 0.121] \\
commit integrators &
  Extraversion &
  242 &
  1.000 &
  [-0.050, 0.084] \\
(core devs) &
  Agreeableness &
  242 &
  1.000 &
  [-0.052, 0.102] \\
 &
  Neuroticism &
  183 &
  1.000 &
 [-0.105, 0.038]\\
  \hline
\end{tabular}\par 
\end{table}

\textit{rq5---What personality traits are associated with the likelihood of becoming a project contributor?}

To answer the fifth research question, \citeauthor{calefato2019apache} built a contribution likelihood model, that is, they fit a logistic regression model to study the associations between the personality traits of developers and the likelihood of having a contribution accepted.
As the response variable, they used a dichotomous yes/no variable indicating whether a developer has authored at least one commit successfully integrated into a project repository. 
They also included a couple of control variables, namely word\_count, a proxy for the extent of communication and social activity of a developer in the community through email messages from which personality traits are extracted, and project\_age, measured as number of days. 
In the replication, we follow the same process described in the original.
The dataset of $N=211$ developers is fairly balanced with respect to the response variable, as 118 developers have at least one commit and 93 have no commits.
Before fitting the model to the new dataset, we first check for the presence of high Pearson correlations between the updated personality predictors; 
We drop \textit{neuroticism} and \textit{extraversion} because they show \revone{a} high correlation ($~.70$) with \textit{conscientiousness} and \textit{agreeableness}, respectively.
Also, the Variance Inflation Factor (VIF) computed on the resulting model reveals no collinearity issues for the retained predictors (all values < 3).
Then, we evaluate the model fit using McFadden's pseudo-\ensuremath{R^2} measure, which describes the proportion of variance in the response variable explained by the model, and the Area Under the ROC curve (AUC), to assess the classification ability of the contribution model as compared to random guessing.

Table~\ref{tab:rep2_rq5_logit} includes the results of the two logistic regression models. 
We observe that the control variable project\_age is statistically significant ($p<0.001$) and has a negative effect in both the original study (-0.420) and the replication (-0.394). 
In the original study, the only statistically significant predictor was openness (54.09, $p< 0.01$), whereas in the replication no personality factor has a significant effect.
Also, in terms of goodness of fit, the model in \citeauthor{calefato2019apache} fit the data slightly better ($R^2=0.397$) than the model in the replication ($R^2=0.270$).
Finally, we replicate the assessment of the model prediction performance computing the AUC. 
As in the original study, we use a stratified sampling technique to split the dataset into training (70\%) and test (30\%) sets.
The AUC performance of the logistic model in the original study is also better than the replication (0.89 vs 0.69). 
Overall, while the results from the original study indicated that higher \textit{openness} scores are associated with better chances for developers to become project contributors, the replication suggests instead that personality traits have no effect.

\begin{table}[tb]
\caption{{Logistic regression model of the contribution likelihood as explained by personality traits in the original study by \fullcite{calefato2019apache} and the replication. Significant results are shown in \textbf{bold} (sig: **$p<0.01$, ***$p<0.001$).} }
\label{tab:rep2_rq5_logit}
\def\arraystretch{1}
\ignorespaces 
\centering 
\begin{tabular}{llrr}
\multicolumn{4}{c}{\textbf{Original study}}\\
\hline 
 & \textbf{Coef. Est.} & \textbf{Std. Error} & \textbf{z-value}\\
\hline 
(Intercept) &
  -29.523 &
  20.175 &
  -1.44\\
project\_age (days) &
  \textbf{-0.420} *** &
  0.113 &
  -3.71\\
log(word\_count) &
  0.199 &
  0.204 &
  0.98\\
openness &
  \textbf{54.092} **&
  23.338 &
  2.32\\
conscientiousness &
  -18.994 &
  26.623 &
  -0.71\\
extraversion &
  -4.652 &
  16.939 &
  -0.27\\
agreeableness &
  18.620 &
  22.525 &
  0.83\\
neuroticism &
  -19.710 &
  16.939 &
  -1.07\\\hline
\multicolumn{4}{l}{N=211, McFadden Pseudo-R\ensuremath{^{2}}=0.397, AUC=0.89} 
  \\\hline
\multicolumn{4}{l}{} 
  \\
\multicolumn{4}{c}{\textbf{Replication}}\\
\hline 
& \textbf{Coef. Est.} & \textbf{Std. Error} & \textbf{z-value}\\
\hline 
(Intercept) &
  1.310 &
  12.867 &
  0.10 \\
project\_age (days) &
  \textbf{-0.366} *** &
  0.080 &
  -4.55 \\
log(word\_count) &
  -0.058 &
  0.146 &
  -0.40 \\
openness &
  -0.532 &
  2.823 &
  -0.19 \\
conscientiousness &
  1.027 &
  3.021 &
  0.34 \\
extraversion &
  0.538 &
  2.505 &
  0.34 \\\hline
\multicolumn{4}{l}{N=211, McFadden Pseudo-R\ensuremath{^{2}}=0.270, AUC=0.69} 
  \\\hline
\end{tabular}\par 
\end{table}

\textit{rq6---What personality traits are associated with a higher number of contributions successfully accepted?}

To answer the last research question, \citeauthor{calefato2019apache} performed a regression analysis to evaluate the association between the personality traits of developers and the number of contributions (i.e., commits) that they got accepted (i.e., merged) into the project repository.
As the independent variables, they used the same personality predictors used in the previous logistic regression analysis.
Regarding the control variables, in addition to the count of words in emails and the age of projects, they added two more (is\_integrator and track\_record) to control for, respectively, core members and long-time contributors.
The dependent variable used is the number of merged commits, i.e., the count of commits authored by a developer that have been successfully merged.
Because the dependent variable takes non-negative integer values only, rather than fitting a linear model, \citeauthor{calefato2019apache} performed a count-data regression analysis, which handles non-negative observations.
Here we follow the same process described in the original work.
Different count data models can be used for estimations, depending on the characteristics of the data. 
Poisson distributions have a strong assumption on equidispersion, that is, the equality of mean and variance of the count-dependent variable.
Alternatively, it is possible to use a negative binomial distribution, a generalization of the Poisson distribution with an additional parameter to accommodate the overdispersion. 
We perform the Likelihood Ratio Test (LRT) of overdispersion and find out that, as in the original study, the negative binomial model (LogLik = -1023, $\chi^2 = 635$, $p < 0.001$) provides a better fit to the data than the Poisson model (LogLik =1340).

Table~\ref{tab:rep2_rq6_count-data-model} shows the results of the count-data regression analysis with the negative binomial models from the two studies. 
We observe that, except for word\_count, all the control variables have a statistically significant effect in both studies. 
Instead, while in the original study none of the five predictors related to personality has a significant effect, in the replication we find that \textit{conscientiousness} is significantly and positively associated with a higher number of integrated commits (0.123, $p < 0.05$). 
Finally, both the model in the original study ($R^2 = 0.115$) and the replication ($R^2 = 0.109$) fit the data marginally.
Overall, while the results from the original study indicated that personality traits do not affect commit productivity, the replication suggests instead that higher \textit{conscientiousness} scores are associated with a higher number of accepted commits. 

\begin{table}[t]
\caption{{Developers productivity model in the original study by \fullcite{calefato2019apache} and the replication. The response is the count of commits successfully merged. The number of observations (commit data) is $N=471$, coming from 211 developers, of whom 118 have made at least one commit. Significant results are shown in \textbf{bold} (sig: * $p<0.05$, **$p<0.01$, ***$p<0.001$).}}
\label{tab:rep2_rq6_count-data-model}
\def\arraystretch{1}
\ignorespaces 
\centering 
\begin{tabular}{llrr}
\multicolumn{4}{c}{\textbf{Original study}}\\
\hline 
  &
  \textbf{Coef. Estimate} &
  \textbf{Std. Error} &
  \textbf{z~value}\\\hline
(Intercept) &
  0.807 &
  0.234 &
  3.43\\
project\_age~(days) & 
  \textbf{-0.068} * &
  0.044 &
  -1.56\\
dev\_is\_integrator=TRUE &    
  \textbf{0.648} ** &
  0.221 &
  2.93\\
dev\_track\_record~(days) &     
  \textbf{0.544} *** &
  0.033 &
  16.21\\
log(word\_count) &    
  0.003 &
  0.030 &
  0.12\\
openness &     
  0.036 &
  0.068 &
  0.53\\
conscientiousness &     
  0.005 &
  0.072 &
  0.08\\
extraversion &
  0.046 &
  0.066 &
  0.71\\
agreeableness &   
  -0.039 &
  0.054 &
  -1.80\\
neuroticism &  
  0.141 &
  0.078 &
  -1.80\\\hline
\multicolumn{4}{p{\dimexpr(.42169999999999995\linewidth-2\tabcolsep)}}{N=471, LogLik=-917, LRT \ensuremath{\chi }\ensuremath{^{2}}=514\mbox{}\protect\newline McFadden Pseudo-R\ensuremath{^{2}}=0.115}\\
\hline 
\multicolumn{4}{l}{}\\
\multicolumn{4}{c}{\textbf{Replication}}\\
\hline 
  &
  \textbf{Coef. Estimate} &
  \textbf{Std. Error} &
  \textbf{z~value}\\\hline
(Intercept) &
   1.169&
   0.175&
   6.67\\
project\_age~(days) & 
   \textbf{-0.087} *&
   0.042&
   -2.06\\
dev\_is\_integrator=TRUE &    
   \textbf{0.474} *&
   0.204&
   2.33\\
dev\_track\_record~(days) &     
   \textbf{0.566} ***&
   0.033&
   17.11\\
log(word\_count) &    
   -0.042&
   0.024&
   -1.76\\
openness &     
   0.004&
   0.054&
    0.07\\
conscientiousness &     
   \textbf{0.123} *&
   0.051&
   2.42\\
extraversion &
   -0.054&
   0.058&
   -0.92\\\hline
\multicolumn{4}{p{\dimexpr(.42169999999999995\linewidth-2\tabcolsep)}}{N=471, LogLik=-1022.984, LRT \ensuremath{\chi }\ensuremath{^{2}}=\mbox{635}\protect\newline McFadden Pseudo-R\ensuremath{^{2}}=0.109}\\
\hline 
\end{tabular}\par 
\end{table}

\subsection{Threats to Validity}
\label{sec:threats-phase2}

Because in the replications we followed the same methodologies presented in the original studies, we have also inherited some of the threats to the validity of those papers, e.g., that the datasets used in \fullcite{iyer2019github} and \fullcite{calefato2019apache} are respectively not representative of GitHub and the Apache ecosystem as a whole.
Also, albeit one may argue that some of the statistics applied in Sect.~\ref{sec:ph2-results} may not be the preferred approach, we applied them to support the comparative aspects of the replication.

\begin{tcolorbox}[standard jigsaw,
	title=Phase 2 -- Summary, opacityback=0]

    The choice of a personality detection tool does affect the validity of previously published results.
    The replication of the first study led to contrasting findings in all the three original research questions aimed to assess the effects of personality on pull-request acceptance.
    When replicating the second study, we were able to obtain consistent findings for only three original research questions out of six.
\end{tcolorbox}


\section{Discussion}
\label{sec:discussion}

\subsubsection*{\revone{The challenges and potential of computational personality detection in \SE\ research}}
\revone{
There has been considerable interest in applying natural language processing (NLP) and computational linguistics to recent \SE\ research.
In particular, prior work on \textit{sentiment analysis} (also referred to as \textit{opinion mining}) has focused on analyzing corpora of technical text, such as emails, commit comments, code-review discussions, and app reviews, to detect the polarity~\cite{calefato2018sentiment,novielli2021offtheshelf}, emotions~\cite{novielli2018gold,calefato2019emtk}, opinions~\cite{lin2019opinionmining,uddin2019opinionmining}, and intentions~\cite{disorbo2015intentionmining,huang2020intentionmining} in software developers' interactions.

One of the reasons for such widespread interest  is that sentiment analysis is a highly restricted NLP problem because, to solve it, tools do not need to fully understand the semantics of each sentence or document but only some aspects of it, i.e., positive or negative sentiments and their target entities or topics~\cite{liu2012saombook}. 
Computational personality detection is also an NLP problem as it touches every aspect of the research field, e.g., co-reference resolution, negation handling, named-entity recognition, and word-sense disambiguation~\cite{cambria2013newavenues}.
However, 
when the NLP aspects are addressed, the related constructs are then used to support the further analyses needed to extract personality profiles, which represents a separate  problem.
The broad scope of the problem arguably explains the poor agreement among the currently available personality detection tools and the negative results of the replications, discussed next in this section.
Also, analogously to sentiment analysis, we expect that computational personality detection in \SE\ will benefit from future breakthroughs and advancements in NLP~\cite{sawant2021nlpbreakthru}.

Furthermore, while sentiment analysis and personality detection are under the same umbrella of \textit{affective computing} research and, therefore, adopt similar technological solutions, the `affective phenomena' they study vary in duration, ranging from short-lived feelings, emotions, and opinions to long-lived, slowly changing personality characteristics~\cite{picard2000affectivecomputing}.
As such, the two research fields  can complement each other also when applied to \SE\ research, with sentiment analysis concentrating on transient feelings related to entities (e.g., others, themselves, objects, and events) and computational personality detection focusing on intrinsic, long-lasting dispositions (of developers). 
For example, previous work on sentiment analysis in \SE\ has also looked into anger~\cite{gachechiladze2017anger} and toxicity~\cite{raman2020toxicity} detection, and identified cases of developers who lashed out at others during technical discussions. 
Computational personality detection could complement this research and tell us if those episodes were extemporaneous or rather the effect of a personal inclination of developers who, despite their technical skills and knowledge, might not be recommended, for example, for tasks such as mentoring newcomers.
Prior research on onboarding has developed  recommender systems that look at developers' social and technical aspects to help newcomers identify mentors in OSS projects~\cite{canfora2012yoda,steinmacher2012recommending}; given that  personality mismatch between mentors and mentees has been identified as one of the social barriers to onboarding~\cite{balali2018newcomers}, a potential follow-up study could  take into account the most relevant traits for the task according to personality theories (e.g., being more agreeable and open to collaboration) to identify candidate developers as suitable for mentoring. 
Another potential scenario of usage concerns code review.
As prior work has shown the impact of human factors in performing such activity~\cite{ruangwan2019codereview}, one might envision the development of a recommender system that assigns developers to code-review tasks also based on their personality profiles, e.g., by preferring those who exhibit a high level of conscientiousness, a trait associated with carrying out tasks precisely and thoroughly.
}

\subsubsection*{Phase 1 -- Performance Assessment}
Because model-based predictions aim to provide an assessment of personality, it is important to establish their convergent validity with self-report scores.
Accordingly, to answer RQ1~(\textit{How do  off-the-shelf personality detection tools perform in the software engineering domain}), in Phase 1 of the study we have built a dataset of emails written by 50 \asf\ developers and compared the predictions of four personality detection tools against the self-ratings collected through a questionnaire.
In addition, we have made further pairwise comparisons between the tool predictions. 

As can be observed from Table~\ref{tab:correlations_results}, the results of the correlation analysis indicate that the tools analyzed neither agree with self-reported personality ratings nor with each other when used in the software engineering domain.
The coefficients are worse than those reported in the literature (see Table~\ref{tab:correlations_comparisons}).
Consistently, as can be observed from Table~\ref{tab:error_results}, the performance accuracy in terms of prediction errors is considerably worse as compared to the results from prior work on automatic personality detection from text (see Table~\ref{tab:error_comparisons}). 
These results should warn the research community about the current limitations when using general-purpose tools for personality prediction in \SE\ research.

\revone{The  disagreements among tools arguable explain why it is hard to synthesize the results from prior work on computational personality detection in \SE\ (see the related work presented next in Sect.~\ref{sec:rel-work}).}
We currently ignore the reasons for these disagreements\revone{, though}. 
On the one hand, a manual error analysis is impracticable in our case.
Even if for simplicity we resorted to using trait predictions based on discrete labels (e.g., low vs. high \textit{openness}), the number of words that the tools need to reliably infer traits is in the hundreds or thousands, too large for us to analyze and reason about the root causes of misclassification.
On the contrary, sentiment analysis \revone{tasks such as polarity detection,} which infers the positive vs. negative polarity conveyed through text, \revone{rely} on machine learning and natural language processing techniques similar to those used for building personality prediction models~\cite{nanli2012sareview} but typically \revone{analyze} input text at the sentence level;
therefore, through manual analysis, researchers on sentiment analysis in \SE\ have identified domain-specific errors limiting model accuracy due to the use of technical words such as \textit{patch} and jargon like \textit{kill a process}, which do not express any valence~\cite{novielli2015challenges}.
On the other hand, the assessed personality detection tools do not explain their outcomes---they are applied in a black-box manner and no information is provided about what steers their model predictions.
The complexity and performance of machine learning models have increased over the years at the expense of interpretability.
Model interpretability is not a monolithic concept and has many facets~\cite{lipton2018interpretability}.
It can be applied at the model level, to refer to algorithmic transparency---a property that is usually very low in deep learning methods whose behavior is notoriously hard to `mentally simulate' by humans; alternatively, it can be applied at the local level, when models offer \textit{post hoc} explanations---like images, text, or examples---for the output generated in response to a single input instance.
Although the need for model interpretability in computational personality detection may not be mandatory because there are no ethical concerns and accountability as in other domains like healthcare and finance, this lack of transparency is nonetheless a drawback that hinders progress in the field.
We argue that the research field on computational personality detection might advance and benefit from the development of transparent prediction models that allow for error analysis.

\subsubsection*{Phase 2 -- Replications}  
Replications play a key role in empirical \SE\ so that the research community can build knowledge about which results or observations hold under which conditions~\cite{shull2008replications}.
To answer RQ2 (\textit{How does the choice of a personality detection tool affect the validity of previous results in \SE\ research?}), in Phase 2  we performed two exact dependent replications in which we kept the same experimental set-up of the original studies while only replacing \pin\ with \liwc\ to infer the developers' personality scores from written documents.

When replicating the first study by \fullcite{iyer2019github}, we have been unable to confirm any of the findings concerning the effect of specific personality traits on the likelihood of merging pull requests in GitHub.
In replicating the second study by \fullcite{calefato2019apache}, we have found consistent results only for three out of six research questions. 
In particular, we have been able to replicate those research questions that failed to find differences in the distributions of median trait scores between subgroups of developers (e.g., core vs. peripheral, high- vs. low-activity). 
Instead, we have failed to replicate the other research questions similar to those reported in \fullcite{iyer2019github}, where a couple of regression models were built to uncover the effects of specific personality traits on the likelihood of becoming a contributor and the productivity level.

As noted by \fullcite{ferguson2012negres}, replicability in science cannot be meaningful without the potential acknowledgment of failed replications.
Negative results may happen when experimental results fail to meet expectations due to a lack of effect rather than  misaligned expectations or a lack of methodological rigor in poorly designed experiments. 
Negative results are uncommon in the literature, even rare in software engineering where only recently there have been specific conference tracks or journals' special issues organized to present such results~\cite{paige2017negres-si}. 
However, negative results are fundamental in software engineering to embrace the nature of experimentation.
In fact, negative results are just as useful as positive ones because, by pointing out what has not worked, they eliminate useless hypotheses and directions, thus redirecting future experimental efforts towards alternative approaches that might pay off~\cite{tichy2000negres}.

Therefore, though we have failed to replicate the experiments from the two studies, we argue that these negative results can be beneficial to enhance the state of the art of computational personality detection in \SE.
\revone{Arguably, the main implication of our findings is} that the validity of previous studies, \revone{including but not limited to the} ones by~\fullcite{iyer2019github} and \fullcite{calefato2019apache}, should be questioned and possibly reassessed. 
However\revone{, this reassessment, as well as future studies on personality detection in \SE, are possible only after developing and testing a reliable, SE-specific personality detection tool.}
\revone{In fact}, the tools used in both the original studies and our replications introduced a threat to validity since all the instruments available for computational personality detection have been trained on non-\SE-specific text documents, such as essays and social media posts.
Hence, the re-assessment should ideally happen using personality detection tools specifically trained on \SE-specific text corpora.
Previous research on sentiment analysis in \SE\ has highlighted the benefits (e.g., the reduced misclassification rate of neutral and positive content as emotionally negative) deriving from the use of tools specifically trained on technical documents retrieved from sources like Stack Overflow and Jira~\cite{calefato2018sentiment}.

Furthermore, in this work, we have used personality detection tools as off-the-shelf components, i.e., without any tuning or training. Recent work on sentiment analysis has shown that the fine-tuning of tools to the \SE\ domain might not be enough to improve accuracy, and that retraining has the potential to adjust the model performance to the shifts in lexical semantics due to different jargon and conventions used in data sources~\cite{lin2018howfar,novielli2021offtheshelf}.
Instead, we observe a trend in recent work on computational personality detection focused on using state-of-the-art, Deep Learning techniques focused more on outperforming baselines (e.g., \cite{mehta2020essay,jiang2020bert,kazameini2020bert,majumder2017deeplearning}) than on releasing reusable, possibly retrainable models that can be transferred to other domains.


\section{Related Work}
\label{sec:rel-work}

In this section, we focus on reviewing previous studies that investigated the Big Five personality model in the \SE\ domain by using tools for automatically extracting personality profiles from communication traces, such as emails, Q\&A posts, and code-review comments.

\fullcite{rigby2007liwc-asf} were the first to automatically analyze the personality traits of developers.
They studied the personality traits of the four top developers of the Apache \texttt{httpd} project against a baseline built by applying \liwc\ on the entire mailing-list corpus. 
They found that two of the developers responsible for the major releases have similar personalities, which are also different from the baseline of all other project members.

\fullcite{licorish2015personalitygsd} combined social network analysis with computational personality detection. They used \liwc\ to analyze the communication traces of 146 practitioners from the IBM Rational Jazz projects involved in global software development activities and found that those who occupy critical roles in knowledge diffusion demonstrate more \textit{openness} to experience.

\fullcite{rastogi2016liwc-gh} used \liwc\ to analyze the personality profiles of nearly 400 GitHub developers. 
They found that those with different levels of contributions have different personality profiles, i.e., those with high or low levels of contributions are more neurotic. 
Also, the personality profiles of most active contributors were found to change across two consecutive years, evolving as more conscientious, more extrovert, and less agreeable.

\fullcite{paruma2016ibmpi} used \pin\ to extract the personality traits from e-mails sent by the committers to six Eclipse projects.
They found three personality clusters: the first personality groups the committers with the highest scores in \textit{extraversion} and \textit{neuroticism};  the second cluster groups the committers with moderate levels of \textit{neuroticism}; the third cluster groups the committers with low values in \textit{neuroticism}.
The three personality clusters are different from those identified by \fullcite{calefato2019apache} after analyzing with the same tool the emails written by the \asf\ developers.

\fullcite{calefato2017trust} investigated the relationship between project success and the propensity to trust, one of the  \textit{agreeableness} facets in the Five-Factor Model. 
They approximated the overall performance of two \asf\ projects with the history of successfully merged pull requests in GitHub.
Using the \liwc-based version of \pin, they analyzed the word usage in pull request comments to extract the developers' agreeableness scores.
The results suggested that the propensity to trust of code reviewers (integrators) is an antecedent of successful pull request integration. 

To the best of our knowledge, this study is the first attempt at replicating results from previous work on computational personality detection in \SE. 
However, one partial exception is the work of \fullcite{bazelli2013liwc-so} who performed a quasi-replication of the study by~\fullcite{rigby2007liwc-asf}.
Specifically, they used \liwc\ to infer the personality of Stack Overflow users from Q\&A posts. 
They found that the top reputed authors on Stack Overflow are more extroverted, as compared to medium- and low-reputed users.
They argued that such a personality profile is consistent with the one observed by \citeauthor{rigby2007liwc-asf} regarding the two top Apache \texttt{httpd} developers.

Overall, the findings from these studies show the existence of different clusters of personalities among developers and that their traits vary with their degree of contribution and reputation, while also changing over short periods.                     
Yet, the negative results of our replications suggest that these results should be also reassessed.


\section{Conclusions}
\label{sec:conclusions}

In this paper, we have studied the impact of the choice of a personality detection tool when conducting \SE\ studies. We have observed a decrease in performance when general-purpose tools are used out of domain as neither they agree with each other nor with the self-reported personality scores.
Also, we have observed that the disagreement among tool predictions can lead to diverging conclusions, making it impossible to replicate previously published results when different personality detection tools are used.
Our results suggest a need for personality detection tools specially targeted for the \SE\ domain.
We hope that sharing the complete replication package---the technical corpus annotated with self-reported personality scores and the experimental workflow scripts---can accelerate the advancement in the field.

\begin{acks}
We are grateful to Filippo Lorè, \textit{Esq}. for his feedback on GDPR compliance.
We also thank our CS students Saverio Telera and Marco Iannotta for their help with the replications.
Part of the computational work has been executed on the IT resources of the ReCaS-Bari data center.
\end{acks}

\bibliographystyle{ACM-Reference-Format}
\bibliography{bibliography}

\newpage
\appendix

\section{Appendix: Replication Package}
\label{appendix:rep-package}

\subsection*{Data Collection and Protection}

The complete replication package is available on Zenodo at \url{https://zenodo.org/record/4679303}.

We are aware of the sensitive nature of the data collected and the privacy risks that come from misusing them.
As such, here we clarify all the measures taken to ensure data privacy and protection during our research work.
Albeit the GDPR took place in May 2018 (i.e., after the data collection conducted in January 2018), we nonetheless made  efforts to comply with the directives already approved by the EU in May 2016.

To ensure that we had control over the data and that they were stored in servers located in the EU, rather than administering the survey through platforms such as Google Forms, we opted for developing in-house an electronic version of the Mini-IPIP personality survey.
As such, the application was hosted on our University cloud infrastructure in Italy.
The application also handled the sending of invitations to participants via email and the automatic removal of such messages after one month.
Both the invitation emails and the landing page of the application identified the research team, clarified the research purpose, and contained a link to the privacy statements briefly summarized next.

In particular, we clarified that: we had retrieved their email address from the public archives of the \asf; there was not going to be any other follow-up email; there was no monetary compensation and study participation was voluntary; the survey responses were needed for research purposes and were going to be stored anonymously; the data and analysis results would be only shared in scientific venues, such as conferences and journals, and presented in an aggregate form, thus making it impossible to identify who participated.
We also clarified that the link to the survey in the invitation email contained a randomly generated id that would match their survey responses to a corpus consisting of some of their emails (i.e., the gold standard) publicly archived by the \asf. 
However, we highlight that (i) at the end of the data collection, we erased all the emails from those who did not answer the survey; (ii) regarding the respondents, by submitting the survey, the system replaced their plain-text email address with the hashed survey id, thus preventing us and anyone else to match an author in the email corpus with their survey responses.

Furthermore, because the \asf\ email archives are publicly available, to further ensure  anonymity and prevent third parties to guess the identity of the survey respondents from the content of their emails, we point out that the gold standard shared in the replication package is built after scrubbing from the email bodies any potentially-sensitive information, such as email and mail addresses, URLs, names, pieces of code, numbers, and stop words (we used three Python libraries, \texttt{clean-text}, \texttt{scrubadub}, and \texttt{NLTK}). 

In conclusion, we are confident that all the measures taken are effective in protecting the privacy and anonymity of the developers who agreed to participate in the study.

\subsection*{Download and Setup}

First, clone the repository and its submodules from GitHub:

\code{git clone $--$recursive https://github.com/collab-uniba/tosem2021-personality-rep-package.git}

Then, before the first execution, run the \code{setup.sh} script.
Also, make sure that the requirements are satisfied, in particular Python 3.8.3+, R 4.0.4+, and Java 1.8+.

\code{bash setup.sh}

\subsection*{Execution Instruction}

It is possible to automatically execute the full experimental workflow by launching the \code{repro.sh} script as follows:

\code{bash repro.sh $--$stage all $--$dataset full}

\revone{For test purposes, instead of supplying the argument \code{full}, it is possible to use the argument \code{test} to work with a small, random subsample of the experimental dataset to shorten the execution time (see the next subsection for more)}:

\code{bash repro.sh $--$stage all $--$dataset test}

It is also possible to execute the two workflow stages independently, using either the full dataset or the test one:

\code{bash repro.sh $--$stage phase1 $--$dataset full}

\code{bash repro.sh $--$stage phase2 $--$dataset test}

\revone{
\subsection*{Execution Times}
The execution of the entire pipeline is quite time-consuming as, depending on the machine specifications, it takes hours---if not days---when working on the full dataset. 
Additional time is also necessary in case one wants to retrain the \tp\ models instead of using the pre-trained ones.

Table~\ref{tab:time} compares the execution times between two machines with different hardware specifications, running the same OS (Ubuntu 20.04). 
We observe that the most recent machine's performance when re-executing the full pipeline on the smaller test dataset is somewhat close to the older machine's (5m17s and 7m35s, respectively).
The newer machine is also faster when it comes to model retraining (6m27s vs. 10m18s).
The largest difference, however, is observed when using the full dataset, which increases the execution time to \tildex{2} days for the more recent configuration and \tildex{3.9} days for the older machine.

}

\begin{table}[hb]
\revonecaption
\revonetable
\small
\newcommand{\cBlueSimple}{\ifdiff\color{blue}\else{}\fi}
\caption{A comparison of execution times on two machines with different hardware specifications running the same OS (Ubuntu LTS 20.4.2).}
\label{tab:time}
\begin{tabular}{llccc}
\hline
\textbf{Year} & \textbf{Hardware specifications}                                                                            & \textbf{\begin{tabular}[c]{@{}c@{}}Pipeline \\ test mode\end{tabular}} & \textbf{\begin{tabular}[c]{@{}c@{}}Pipeline\\ full mode\end{tabular}} & \textbf{\begin{tabular}[c]{@{}c@{}}Model\\ retraining\end{tabular}} \\ \hline
2018       & \begin{tabular}[c]{@{}l@{}}CPU Intel i7-7700 (Kaby Lake),\\ 8 cores @ 3.60GHz, 16GB RAM\end{tabular}        & 5m17s                                                                  &  \tildex{2d}                                                                     & 6m27s                                                                                 \\ \hline
2011        & \begin{tabular}[c]{@{}l@{}}CPU Intel Xeon E312xx (Sandy Bridge),\\ 8 cores @ 2.00GHz, 16GB RAM\end{tabular} & 7m35s                                                                  &   \tildex{3.9d}                                                                    & 10m18s                                                                                \\ \hline
\end{tabular}
\end{table}

\newpage
\section{Appendix: Phase 1 -- Additional Material}
\label{appendix:additional-material-ph1}

\begin{table}[!ht]
\caption{Descriptive statistics per trait for the distributions of scores in the gold standard and inferred by the tools.}
\label{tab:desc-stats}
\begin{tabular}{lccccc}
\hline
\textbf{Gold standard} & \textbf{Min} & \textbf{Max} & \textbf{Mean} & \textbf{Median} & \textbf{SD} \\ \hline 
Openness          & 2.75          & 5.00         & 4.31          & 4.50            & 0.63        \\
Conscientiousness & 1.75          & 5.00         & 3.71          & 3.75            & 0.76        \\
Extraversion      & 1.00          & 5.00         & 2.76          & 2.75            & 0.83        \\
Agreeableness     & 1.25          & 5.00         & 3.73          & 4.00            & 0.81        \\
Neuroticism       & 1.00          & 5.00         & 2.77          & 2.75            & 0.93        \\ \hline
\textbf{\liwc}         & \textbf{Min} & \textbf{Max} & \textbf{Mean} & \textbf{Median} & \textbf{SD} \\ \hline 
Openness          & 0.97          & 2.35         & 1.67          & 1.69            & 0.26        \\
Conscientiousness & 1.27          & 2.83         & 2.52          & 2.51            & 0.24        \\
Extraversion      & 2.12          & 3.46         & 3.09          & 3.09            & 0.17        \\
Agreeableness     & 3.24          & 4.30         & 3.60          & 3.58            & 0.17        \\
Neuroticism       & 3.34          & 5.00         & 3.81          & 3.80            & 0.24        \\ \hline
\textbf{\pin}          & \textbf{Min} & \textbf{Max} & \textbf{Mean} & \textbf{Median} & \textbf{SD} \\ \hline 
Openness          & 4.04          & 4.55         & 4.23          & 4.24            & 0.09        \\
Conscientiousness & 3.10          & 3.72         & 3.47          & 3.48            & 0.13        \\
Extraversion      & 2.70          & 3.30         & 2.98          & 2.96            & 0.11        \\
Agreeableness     & 3.42          & 3.92         & 3.61          & 3.59            & 0.11        \\
Neuroticism       & 2.52          & 3.24         & 3.03          & 3.05            & 0.13        \\ \hline
\textbf{\pr}      & \textbf{Min}  & \textbf{Max} & \textbf{Mean} & \textbf{Median} & \textbf{SD} \\ \hline 
Openness          & 2.80          & 4.07         & 3.45          & 3.43            & 0.31        \\
Conscientiousness & 3.00          & 4.07         & 3.51          & 3.52            & 0.23        \\
Extraversion      & 2.33          & 4.17         & 3.52          & 3.53            & 0.29        \\
Agreeableness     & 2.93          & 4.02         & 3.43          & 3.44            & 0.20        \\
Neuroticism       & 2.47          & 3.68         & 2.98          & 2.92            & 0.33        \\ \hline
\textbf{\tp}      & \textbf{Min}  & \textbf{Max} & \textbf{Mean} & \textbf{Median} & \textbf{SD} \\ \hline 
Openness          & 4.13          & 4.14         & 4.13          & 4.13            & 0.01        \\
Conscientiousness & 3.51          & 3.52         & 3.51          & 3.51            & 0.01        \\
Extraversion      & 3.34          & 3.36         & 3.35          & 3.35            & 0.01        \\
Agreeableness     & 3.62          & 3.63         & 3.63          & 3.63            & 0.01        \\
Neuroticism       & 2.60          & 2.60         & 2.60          & 2.60            & 0.00        \\ \hline
\end{tabular}
\end{table}

\begin{figure}[!ht]
\centerline{
\includegraphics[width=0.7\textwidth]{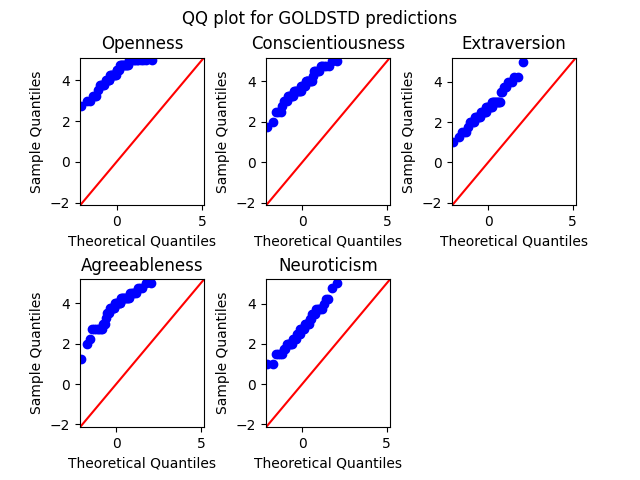}}
\caption{Q-Q plot of the trait distributions from the gold standard.}
\label{fig:qqplot-gs}
\end{figure}

\begin{figure}[!ht]
\centerline{
\includegraphics[width=0.7\textwidth]{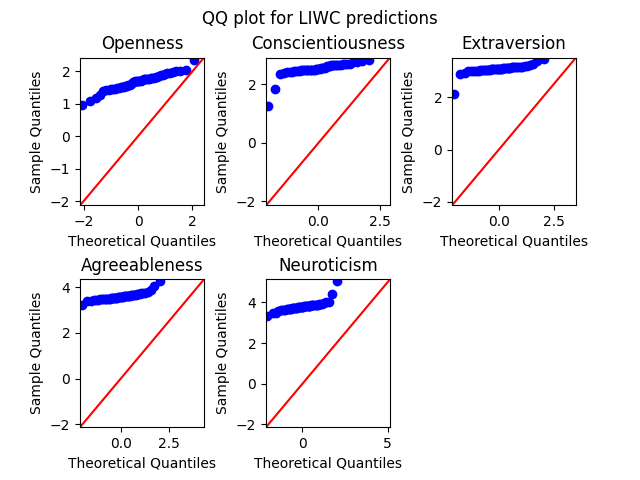}}
\caption{Q-Q plot of the trait distributions from \liwc.}
\label{fig:qqplot-liwc}
\end{figure}

\begin{figure}[!ht]
\centerline{
\includegraphics[width=0.7\textwidth]{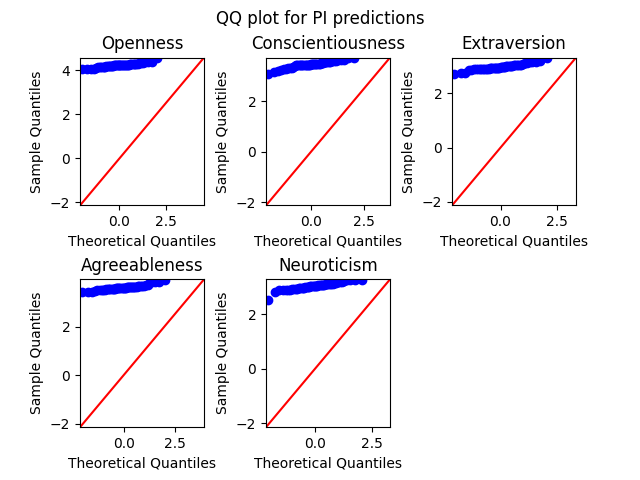}}
\caption{Q-Q plot of the trait distributions from \pin.}
\label{fig:qqplot-pi}
\end{figure}

\begin{figure}[!ht]
\centerline{
\includegraphics[width=0.7\textwidth]{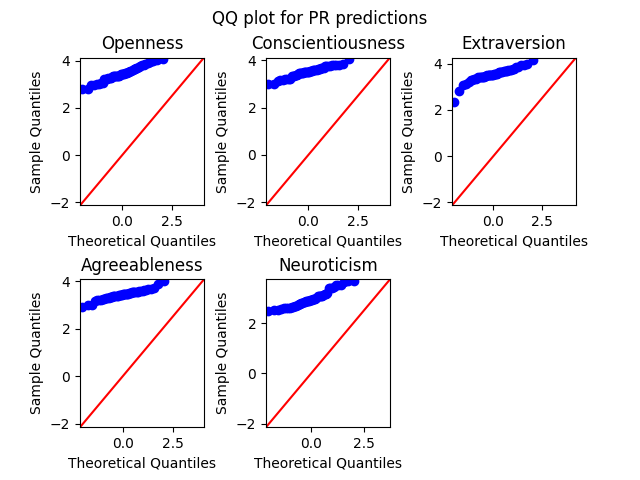}}
\caption{Q-Q plot of the trait distributions from \pr.}
\label{fig:qqplot-pr}
\end{figure}

\begin{figure}[!ht]
\centerline{
\includegraphics[width=0.7\textwidth]{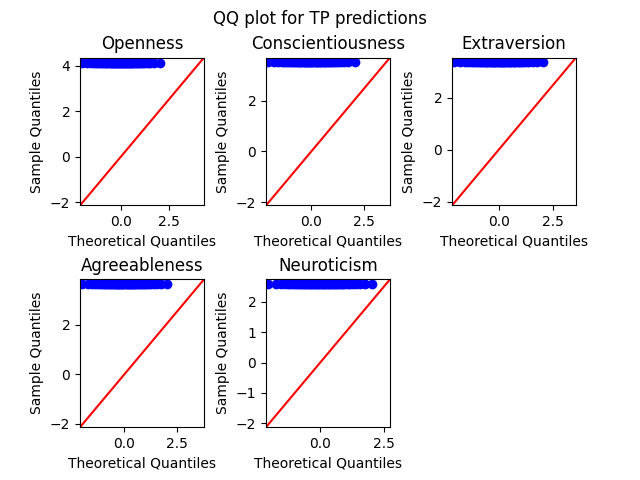}}
\caption{Q-Q plot of the trait distributions from \tp.}
\label{fig:qqplot-tp}
\end{figure}

\clearpage

\section{Appendix: Phase 2 -- Additional material}
\label{appendix:additional-material-ph2}

\begin{figure}[!ht]
\centering
    \subfloat[Original study \label{fig:rq1_cluster_screeplot_orig}]{%
       \includegraphics[width=0.5\textwidth]{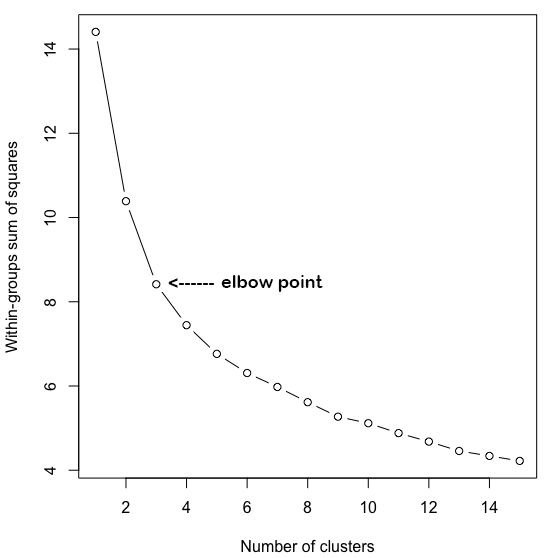}
     }
    \subfloat[Replication\label{fig:rq1_cluster_screeplot_repl}]{%
       \includegraphics[width=0.5\textwidth]{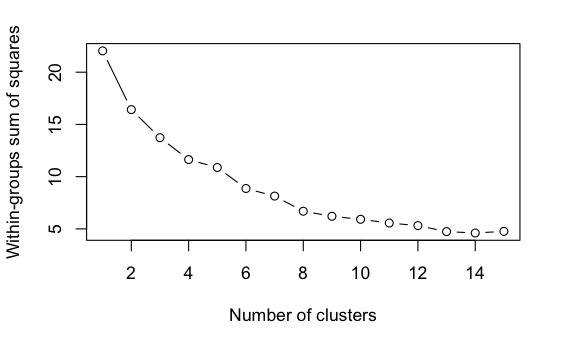}
     }
\caption{Plot of within-group heterogeneity against the number of \textit{k}-means clusters in the original study by~\fullcite{calefato2019apache} (a) and the replication (b).}
\label{fig:rep2_rq1_cluster_screeplots}
\end{figure}

\begin{figure}[!ht]
\centering
    \subfloat[Original study \label{fig:rep2_rq1_archetypes_screeplot_orig}]{%
       \includegraphics[width=0.5\textwidth]{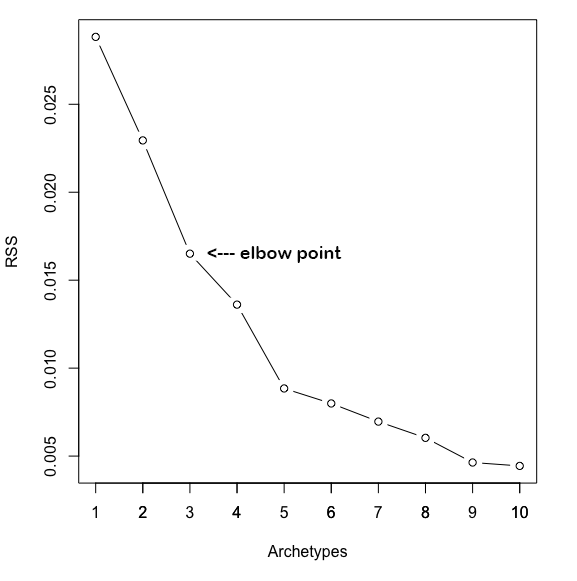}
     }
    \subfloat[Replication\label{fig:rep2_rq1_archetypes_screeplot_repl}]{%
       \includegraphics[width=0.5\textwidth]{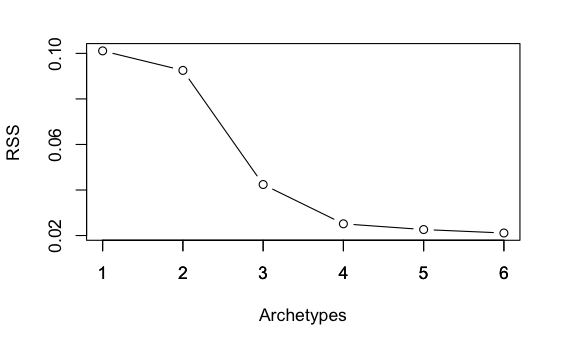}
     }
\caption{Scree plot of the residual sum of squares against the number of archetypes in the original study by~\fullcite{calefato2019apache} (a) and the replication (b).}
\label{fig:rep2_rq1_archetypes_screeplots}
\end{figure}

\begin{figure}[!ht]
\centering
    \subfloat[Original study \label{fig:rep2_rq3_boxplot_orig}]{%
       \includegraphics[width=0.85\textwidth]{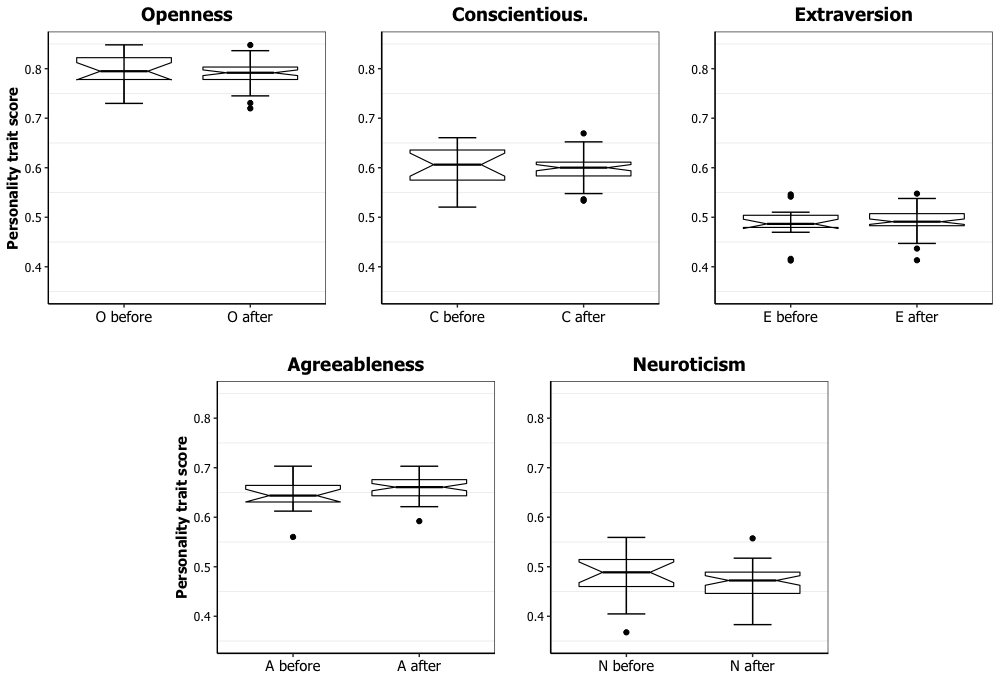}
     }\hfill
    \subfloat[Replication\label{fig:rep2_rq3_boxplot_repl}]{%
       \includegraphics[width=0.85\textwidth]{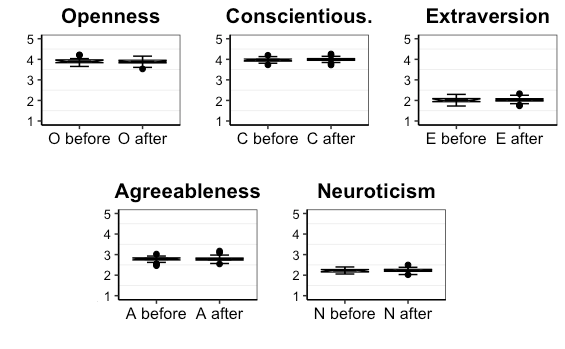}
     }
\caption{Differences in the personality traits of the developers before and after becoming core team members in the original study by~\fullcite{calefato2019apache} (a) and the replication (b).}
\label{fig:rep2_rq3_boxplot_membership}
\end{figure}

\end{document}